\documentclass[aps,prc,twocolumn,showpacs,reprint]{revtex4-1}
\usepackage{ulem,color,graphics}
\usepackage{amssymb,amsmath,textcomp}
\usepackage{graphicx} 
\usepackage{subfigure} 

\usepackage{dcolumn}
\usepackage{times} 
\graphicspath{{04-figuras/}} 
\usepackage{comment}
\usepackage{bm}
    \newcommand{\ket}[1]{\left | #1 \right >}
    \newcommand{\bra}[1]{\left < #1 \right |}
    \newcommand{\cav}{g_{s,\tau}^{s',\tau'}}
    \def\etal{{\it et al. }}
    \usepackage{multirow, array}				
\newcolumntype{L}[1]{>{\raggedright\let\newline\\\arraybackslash\hspace{0pt}}m{#1}}
\newcolumntype{C}[1]{>{\centering\let\newline\\\arraybackslash\hspace{0pt}}m{#1}}
\newcolumntype{R}[1]{>{\raggedleft\let\newline\\\arraybackslash\hspace{0pt}}m{#1}}

\begin{document}


\title{Nuclear medium effects in muonic neutrino interaction
with energies from 0.2 GeV to 1.5 GeV}



\author{D. Vargas}
\author{A.R. Samana}
\email[]{arsamana@uesc.br}
\author{F.G. Velasco}
\author{O.R. Hoyos}
\author{F. Guzm\'an}
\affiliation{Universidade Estadual de Santa Cruz - UESC,
Rodovia Jorge Amado km 16, Ilh\'eus,  45662-900, Brasil}

\author{J.L. Bernal-Castillo}
\author{E. Andrade-II}
\author{R. Perez}
\author{A. Deppman}
\affiliation{Instituto de F\'isica da Universidade de S\~ao Paulo-IFUSP,
Rua do Mat\~ao, Travessa R, 187, S\~ao Paulo, 05508-090, Brasil}

\author{C.A. Barbero}
\author{A.E. Mariano}
\affiliation{Instituto de F\'isica La Plata-CONICET, 49 y 115, La Plata, CP 1900, Argentina.}

\date{\today}

\begin{abstract}
Nuclear reactions induced by muon neutrino with energies in the range from 0.2 to 1.5 GeV
in Monte Carlo calculations framework in the intra-nuclear cascade model are studied.
This study was done by comparison between the available experimental data and theoretical
values of total cross section, and the energy distribution of emitted lepton energy in the
reaction muon neutrino-nucleus, using the
targets: $^{12}$C, $^{16}$O, $^{27}${Al}, $^{40}${Ar}, $^{56}${Fe} and $^{208}${Pb}.
A phenomenological toy model of primary neutrino-nucleon interaction gives a
good agreement of our
theoretical inclusive neutrino nucleus cross 
in comparison with
the available experimental data.
Some interesting results on the behavior of the cross section as function of 1p-1n and
higher contributions are also sketched. The previous results on the percentage of fake events
related in available experiments in $^{12}$C were expanded for the set studied nuclei.
With the increase of mass target, the nuclear effects in the cross sections
were observed
along with
the importance to take into account fake events in the reactions.
\end{abstract}


\maketitle

\section{Introduction}
The investigation of neutrino-nucleus interaction is a field that gained an increasing
relevance in recent years allowing studies on neutrino oscillation and  the neutrino
massiveness \cite{Gall16}.
Moreover, neutrino-nucleus interaction plays a key role
in Astrophysics issues such as supernova dynamics \cite{Lang16}.

From the experimental side, a special difficulty faced in the study of neutrino
interactions is the fact that the neutrino energy is unknown, being described by
broad energy distributions.
This problem prevents the extraction of information
concerning to essential characteristics of neutrinos \cite{Gall16}, which requires
their reconstruction fluxes from final states measurements. Furthermore, final states are strongly dependent on nuclear properties and nuclear effects.

Within the treatment of weak interaction in the nuclear medium appear complex
processes due to the effects of nuclear structure and interactions between the
various nucleons.
There are several theoretical models in the description of neutrino-nucleus/nucleon
cross sections.
A few of them
have been implemented in computer numerical
codes to simulate the interactions in several neutrino experiments in progress.
Many of the used formalisms do not have a specific name, and here we will distinguish
them with the name of the numerical code
such as GENIE ~\cite{Andr09}, NUANCE~ \cite{Casp02},
between others. Further, there is another, not yet implemented
theoretical formalism called Consistent Isobar Model-CIM ~\cite{Barb13,Mari11}.

Many of these codes are using Monte Carlo (MC) procedures
to simulate the reactions in the nuclear cascade.
Some important notes are claimed by Ref.~\cite{Gall16}: (i) presently available
generators all rely on free-particle MC cascade simulations that are applicable
at very high energies with limited applicability in the description of relatively
low energy with Final States Interactions (FSI) inside the target nuclei;
(ii) it is neglected the binding in nuclei and;
(iii) some generators are working
with outdated nuclear physics and there is not an internal consistency
between the different reaction channels.

Another important task that the simulation program and event generator
should take into account is the elimination of those
so-called fake events, from secondary
interactions that introduce noise in the main channel. These secondary interactions
are into a more
fundamental level of neutrino-nucleon interaction theory, requiring a
deep understanding of this interaction.
In many neutrino experiments
are emitted neutrinos
by secondary decays of pions and kaons, usually produced
in high energy proton-nucleon/nucleus collisions. For example, in the
K2K (Kamioka to Kamioka) \cite{K2K06,K2K08} experiment a proton beam of
12.9 GeV collides against Al.
In the MiniBooNE experiment (Mini Booster Neutrino Experiment)
\cite{MiniBooNE10,MiniBooNE11},
a proton beam of 8.9 GeV collides against Be, forming the so-called long-range beam
Long Base Line (LBL). The beams produced in LBL range, from hundreds of MeV to
several GeV, are detected hundreds of kilometers away. In this energy range,
the dominant contribution to the neutrino-nucleus cross section comes from
reactions with charged current (CC) in the channels: quasi-elastic (CCqe) and
resonance (CCres) production.
There are currently several LBL type experiments
in progress, designed to determine the differences between the masses of different
kinds of neutrinos and oscillation parameters.
In this work we do not analyze
the effect on neutral current on the target nuclei here studied, due
that the CCqe scattering is the dominant neutrino
interaction process for $\nu_\mu$ and ${\bar \nu}_\mu$
colliding with a nuclear target  when the
neutrino  energies are on the order 1 GeV \cite{Wilkinson16},
On the other hand,
Ericson {\it et. al}~\cite{Eric2016} have
called the attention in the sense
that  $\nu_\mu$ neutral current
could be necessary
to solve the MiniBooNE low-energy anomaly.

The CCqe process
\begin{align}
\begin{split}
 &\nu_{l}+n\rightarrow l^{-}+p, \\
 &\bar{\nu_{l}}+p\rightarrow l^{+}+n,
 \end{split}
\end{align}
represents the simplest form of neutrino-nucleon (antineutrino-nucleon) interaction,
where the weak charged current induces a transition of neutrino (antineutrino) into
its corresponding lepton charged $l^{-}$ ($l^{+}$), that results
in
the signal of an event. The FSI may lead to more than one ejected nucleon
, plus a lepton
, and resonances produced by absorp
tion of emitted pions can also lead to more
ejected nucleons. These last two contributions affect the reconstruction of energy
and production of quasi-elastic fake events. Many experiments try to reduce these
uncertainties using a near detector and
implementing some correlation
with the far main detector.
Nevertheless,
there are no previous studies on how the event
generator manages these fake events other than the works of Lalakulich
and Mosel~\cite{Mos10,Lalakulich2013}
and alternatively, Ericson {\it et. al}~\cite{Eric2016}
in the quasi-elastic reaction of  $\nu_\mu- ^{12}$C.

In the present paper, we show recent developments on the inclusion of neutrino-nuclear
interaction in the CRISP
(\textbf{C}ollaboration \textbf{R}io-\textbf{I}lh\'eus-\textbf{S}\~ao \textbf{\textbf{P}}aulo)
model \cite{Deppman2004}. CRISP is a nuclear reaction model based on Quantum Dynamics (QD) and Monte Carlo (MC) methods and has being developed for the last two
decades~\cite{Goncalves1997,Deppman2001,Deppman2004,Andrade2011,Israel2014,Andrade2015}.
CRISP provides reliable descriptions of many-body interactions for photons and electrons,
for protons and neutrons, and has being applied to study reactions in nuclei from $^{12}$C
to $^{240}$Am. The incident particles can have energies from 50 MeV up to tens of GeV, and
many aspects of nuclear reaction can be investigated,
such as specific cross sections, particle multiplicity, and particle spectra,
between others. Additionally, the CRISP model has been employed
for the investigation of electron scattering~\cite{Likh03}, meson production in
nuclei~\cite{Israel2014}, ultra-peripheral collisions at LHC energies~\cite{Andrade2015},
and  $\Lambda$ non-mesonic decay in the nuclear
medium~\cite{Gonz11,Israel2011b}
using the smallest numbers of possible free parameters. CRISP has not been used before
to study neutrino-nucleus interaction, then this work is the first study to this issue.

Further, it is a useful tool
to study nuclear effects on different nuclear
reactions, which is not the usual case for codes built as event generators, where many
parameters must be adjusted for specific reactions.
In this paper, we focus on the nuclear
effects in neutrino-nucleus reaction. For this purpose, we first include a simple
toy model of the primary neutrino-nucleon interaction in the CRISP
code, and then analyze
how the nuclear effect modifies the different observables.

The paper is organized as follows: in section \ref{TheorModels} we describe briefly
the CRISP model and introduce a simple toy model for the neutrino-nucleon interaction
that was coupled to CRISP code. In section \ref{ResAna}, the results are presented
and discussed. Finally,  in section \ref{Conc} we show our conclusions and final
remarks.

\section{Theoretical model \label{TheorModels}}
The study of nuclear reactions must consider all relevant effects due to the nuclear
medium.
In this paper, we used the CRISP model for the calculation of nuclear reactions.
The CRISP code was developed to describe the most relevant nuclear processes realistically.
In the following, are presented the most important aspects regarding the nuclear medium.

\subsection{CRISP \label{CRISPmodel}}
QD method and MC method \cite{Deppman2004} are used
in the CRISP model  to describe the nuclear processes that take place during a nuclear
reaction. In CRISP code, the target is constructed as a Fermi gas
where the Fermi energies
for protons and neutrons, respectively, are
\begin{align}
 \begin{split}
E_{F}^{(p)}&=\frac{1}{2\: m_{0}}\: (3\pi^{2})^{2/3}\: \bigg (\frac{Z}{L^{3}}\bigg )^{2/3},\\
E_{F}^{(n)}&=\frac{1}{2\: m_{0}}\: (3\pi^{2})^{2/3}\: \bigg (\frac{A-Z}{L^{3}}\bigg )^{2/3}.
 \end{split}
\end{align}
\noindent
where, $L^{3} = \frac{4}{3}\: \pi {r_{0}}^{3}\: A$,
is the nuclear volume, with $r_0=1.18$ fm, and $m_0$ is the rest nucleon mass.
The ground state from the momentum space
is always generated, including the degrees of freedom related to spin.
The respective
Fermi momenta for protons and neutrons
are given by
\begin{align}
 \begin{split}
k_{F}^{(p)}&=\sqrt{E_{F}^{(p)}(E_{F}^{(p)}+2m_{0})},\\
k_{F}^{(n)}&=\sqrt{E_{F}^{(n)}(E_{F}^{(n)}+2m_{0})}.
 \end{split}
\end{align}

The momentum space is divided into cells of width $\Delta p$ calculated as
\begin{eqnarray}
\Delta p=\frac{k_{F}}{N_{l}},
\end{eqnarray}
where $N_{l}$ represents
the number of levels in the Fermi gas.
All nucleons are evenly distributed inside the nuclear volume.

The nuclear reaction in the CRISP model is considered as a two-step calculation process.
The first one is the intranuclear cascade, described by the
Monte Carlo MultiCollisional (MCMC) model
\cite{Koda82}. The second step is the
evaporation-fission competition, described by
Monte Carlo Evaporation-Fission (MCEF)
model \cite{Depp02,Depp03}. The emphasis of this work is on the intranuclear cascade
step since the particles of interest (muon, muon neutrinos, and pions) are emitted only
during this step. For the sake of completeness, it must be mentioned that in the
evaporation-fission part the Weisskopf's model is used to describe the nuclear
de-excitation process by successive evaporation of nucleons or by nuclear fission
~\cite{Depp03,Deppman2015,Andrade2011,Velasco2016}. In the case of fission,
the fragments are generated following the Random Neck Rupture Model (RNRM)~\cite{Brosa1990}
with symmetric, asymmetric and super-asymmetric channels
~\cite{Deppman2013b,Deppman2013c}
for the fragments formation.
Besides, we include the evaporation of hot fission fragments.

In the intranuclear cascade step, binary interactions only can occur.
The multicollisional
approach implies that all nucleons move simultaneously~\cite{Koda82}. Such an approach
makes it natural to check dynamical aspects such as changes in the nuclear density and
the evolution of the occupancy levels of the Fermi gas~\cite{Deppman2004,Rodrigues2004}.
The Fermi motion of nucleons, also a result of this approach, modifies the nuclear cross
sections, especially near the threshold of the interaction. The ordered sequence of
collisions considers the probability of interaction with all particles, based on their
respective cross sections.

The intranuclear cascade starts with the primary collision
when the incident particle interacts on
the surface a nucleon of the system or more internally in the nucleus.
As a result,  secondary particles are produced which have relatively high energy compared
to the energy of the others nucleons in the nuclear medium. These particles are called
cascade particles. The secondary particles propagate inside the nucleus and can interact
with other particles,
or they can be emitted when reach the nuclear surface just as their
kinetic energy is higher than the nuclear potential or, be reflected
continuing their propagation in the nucleus.
The nuclear potential is a square well such that
\begin{eqnarray}
V_{0}=E_{F}+B,
\end{eqnarray}
where $B$ is the binding energy, $\sim 8$ MeV. The CRISP model also considers
the effect of tunneling of charged particles through the Coulomb barrier.

The cascade is completed when there is no resonance yet to decay or hadrons with kinetic
energy greater than the nuclear potential. After this condition is satisfied, the remaining
excitation energy is evenly distributed between the nucleons in a process known as
thermalization. The main characteristics of the nucleus
does not change at this stage, so that its atomic number, mass number,
and excitation energy remain the same ones until the
end of the process ~\cite{Deppman2004,Rodrigues2004}.

Another fundamental characteristic of CRISP is the strict verification of the Pauli
exclusion principle~\cite{Deppman2004}, possible thanks, both to the application of
the Fermi gas model and the multicollisional approach which
to be known enables the 4-vectors of
all nucleons  at each step of intranuclear cascade.

\subsection{Implementation of the muon neutrino channel as an event generator
of the intranuclear cascade}
\subsubsection{Primary interaction}
The energy range of the muonic neutrino in this
paper is $0,2$ - $1,5$ GeV.
The most important channels in this energy range are the quasi-elastic scattering
and the resonance production.
The formation of the $\Delta$ (1232) dominates
the resonance production, which subsequently decays into a pion and a nucleon.
The first step to study the nuclear effects in neutrino-nucleus interaction is to incorporate  the primary neutrino-nucleon interaction in the CRISP model. In this way we are considering the neutrino-nucleus interaction as an incoherent sum of
the contributions from all nucleons inside the nucleus.
Besides,
CRISP calculates many nuclear effects as those due to the Fermi motion and to the
antisymmetrization  of the nuclear wave-functions, as described in
subsection~\ref{CRISPmodel}, as well as all the possible particle-hole states formed due to final state interactions (labeled  as npnh events).
In the present work,  we will not be considered
a possible coherent contributions.

Here, the primary neutrino-nucleon interaction is formulated through a
toy model where are exactly considered
kinematic and isospin aspects of the interaction.

Due to the limitations of the CRISP model, where the nuclear states
are described as a Fermi gas,
the angular momentum is not a conserved quantity. For these  reasons,
all our calculations are averaged on the spin states, and an
important consequence is the fact that we will not be able
to describe angular distributions correctly. In the following, the model
for the primary interaction is referred
as Kinematic Model (KM).

In our model, the neutrino is supposed to interact with a single
nucleon, and since our goal is to analyze the interaction near the
threshold region, we consider two interaction
channels, namely, the quasi-elastic  and the $\Delta-$resonance
formation,
described below.

\begin{table*}[!ht]
\centering
\caption{ Correspondence between the different labels
and relevant parameters
for each primary interaction used in this work.
The $A_n$ coefficients are in units of $10^{-38}$~cm$^{2}$. The
coefficients $\sigma_0$ are dimensionless.
}
\label{canais}
\begin{ruledtabular}
\begin{tabular}{ c | c | c | c }
Process & Channel & Parameters & Label \\
\hline
CCqe & $\nu_{\mu}+n\rightarrow \mu^{-}+p$
     & $A_{0}=( 2.77\:\pm\:1.30)~\times 10^{-2}$ \hspace{1cm}~~& A \\
    && $A_{1}=( 1.07\:\pm\:0.48)~~$ \hspace{2cm}~~~& \\
	&& $A_{2}=(-1.01\:\pm\:0.31)~\times 10^{-1} $ \hspace{1cm}~& \\
	&& $A_{3}=(-2.45\:\pm\:0.25)~\times 10^{-1} $ \hspace{1cm}~& \\
	&& $A_{4}=( 7.32\:\pm\:1.30)~\times 10^{-2} $ \hspace{1cm}~~~~~& \\
\hline
	&& $\sigma_0$ & \\
CCres&$\nu_{\mu}+p\rightarrow \mu^{-}+\Delta^{++}\rightarrow \mu^{-}+\pi^{+}+p$ ~~~~~~~~& $0.66\:\pm\:0.12$ & B \\
     &$\nu_{\mu}+n\rightarrow \mu^{-}+\Delta^{+}\rightarrow \mu^{-}+\pi^{+}+n$ ~~~~~~~~& $0,26\:\pm\:0,07$ & C \\
     &$\nu_{\mu}+n\rightarrow \mu^{-}+\Delta^{+}\rightarrow \mu^{-}+\pi^{0}+p$ ~~~~~~~~& $0,25\:\pm\:0,07$ & D \\
	    \end{tabular}
	  \end{ruledtabular}
   \end{table*}	

\subsubsection{Quasi-elastic channel}
The CCqe channel corresponds to a quasi-elastic interaction
between neutrino and nucleon
where the charged  current induces isospin modification of the nucleon.
In the process,
the neutrino is absorbed and a muon is produced.
This process is indicated
in Table \ref{canais}.

The empirical formula that gives the CCqe cross
section per nucleon is
\begin{eqnarray}
\sigma^{(CCqe)}= \sum_{n=0}^4 A_n E^n_{\nu}.
\label{ajuste1}
\end{eqnarray}
We implemented in the code
the Equation (\ref{ajuste1}), being
the best fit polynomial of order 4th for experimental deuteron
experimental data in the range of interest. The coefficients $A_n$ are
shown in Table I.

\subsubsection{Resonant channel}
In the initial state, we have a nucleon $N$ with momentum $p_{N}$ a
nd a neutrino $\nu$ with momentum $p_{\nu}$. This state is
represented by $\ket {N,\nu}_{s,\tau}$, where $s,\tau$
are the total spin and isospin of the system neutrino-nucleon.
Let be $\cav$ the coupling constant for the neutrino-nucleon
vertex, and $s',\tau'$ the spin and isospin of the
final state.

We are interested in resonant states, so we project the final
states onto the resonant states, resulting in
\begin{multline}
\ket{{\Psi}_{\Delta}}= A \int d^{4} p_{\Delta} \int d^{4}
p_l \sum_{ss'}
\ket{\Delta, l} \times
\\
\bra{\Delta,l}_{s', \tau'} \cav \ket{N,\nu}_{s, \tau}
\times \delta^{4}(p_{\Delta}-p_{l}-p_{N}-p_{\nu}),
\end{multline}
with $A$ being a normalization constant, $p_{\Delta}$ and $p_l$ are,
respectively the resonance and lepton momentum. We sum over all
possible spin configuration and
\begin{align}
\begin{split}
\ket{\Delta,l}_{\tau'} &= \sum_{s'} \ket{\Delta,l}_{s',\tau'},\\
\ket{N,\nu}_{\tau} &= \sum_{s} \ket{N,\nu}_{s,\tau},
\end{split}
\end{align}
then
\begin{multline}
\ket{{\Psi}_{\Delta}}= A \int \frac{1}{E_{l}}d^{3}p_{l}
\int dm_{\Delta} {\ket{\Delta,l}}_{\tau'} \:{}_{\tau'}
{\bra{\Delta,l}} g_{\tau}^{\tau'}
\ket{N,\nu }_{\tau},
\label{9}\end{multline}
where was performed the integration
on $d^{3}p_{\Delta}$, then
$p_{\Delta}=p_{N}+p_{\nu}+p_{l}$, and the coupling constant
for the neutrino-nucleon vertex
was renamed for simplicity of notation, $\cav \equiv g_{\tau}^{\tau'}$.
This integration on $m_{\Delta}$ is equivalent
to the integration of the total $\Delta$ energy, $E_{\Delta}$.

The resonant state propagates through the Hamiltonian
$H_{\Delta,l}=H_{\Delta}+H_{l}$. From the Lippman-Schwinger
equation in first order approximation, we have
\begin{align}
\begin{split}
\ket{{\Psi}_{\Delta}}=& A \int \frac{1}{E_{l}}d^{3}p_{l}
\frac{g^{N}_{\Delta}}{\big(E^{*}-H_{\Delta} + i \frac{\Gamma}{2} \big)}
\ket{\Delta,l}_{\tau'}\\
&\times\: {}_{\tau'}\bra{\Delta,l} g^{\tau'}_{\tau}
\ket{N,\nu}_{\tau},
\label{10}
\end{split}
\end{align}
with $E^{*} = E-E_{l}$, where $E_{l}$ is the lepton energy,
$E$ is the energy of the neutrino-nucleon system and, $\Gamma$
is the half-width at half maximum of
the curve. In the above equation, we assume that the lepton
Hamiltonian $H_{l}$ corresponds to the free lepton,
$H_{\Delta}$ is the free resonance Hamiltonian,
and $g^{N}_{\Delta}$ is the resonance-nucleon coupling.

The final states are formed by a final nucleon $N'$,
a lepton and a meson. So,
we project the state in the above equation on states of the
form $\ket{N', m, l }$, leading
\begin{align}
\begin{split}
\ket{{\Psi}_{f}}=& A \int \frac{1}{E_{l}}d^{3}p_{l} \int d^{4} p_{N'}
\int d^{4} p_{m} \ket{N',m,l}_{\tau''} \\
&\times \:{}_{\tau''}\bra{N',m,l} \frac{g^{N}_{\Delta}}{\big(E^{*}
-H_{\Delta}+i\frac{\Gamma}{2} \big)} \ket{\Delta,l}_{\tau'} \\
&\times {}_{\tau'}
\bra{\Delta,l} g^{\tau'}_{\tau} \ket{N,\nu}_{\tau}
\delta^4 (p_{N'}+p_{m}-p_{\Delta}),
\label{11}
\end{split}
\end{align}
which can be integrated with respect
$p_{N'}$, resulting
\begin{align}
\begin{split}
\ket{{\Psi}_{f}}=& A \int \frac{1}{E_{l}}d^{3}p_{l}
\int \frac{1}{E_{m}}d^{3}p_{m}
\ket{N',m,l}_{\tau'} \\
&\times \:{}_{\tau'}\bra{N',m,l} \frac{g^{N}_{\Delta}}
{\big(E^{*}-H_{\Delta}+i \frac{\Gamma}{2} \big)} \ket{\Delta,l}_{\tau'} \\
&\times {}_{\tau'}\bra{\Delta,l} g^{\tau'}_{\tau}
\ket{N,\nu}_{\tau},
\label{12}
\end{split}\end{align}
with $p_{N'}=p_{\Delta}-p_{m}$. Then, the transition probability reads
\begin{align}\begin{split}
d\sigma_0(\tau, \tau')=\left|{\Psi}_{f} \right|^{2}&=
A \frac{\sigma_0(\tau, \tau')} {\big[ (E^{*}-H_{\Delta})^2+(\Gamma/2)^2 \big]} d^{3}p_{l} \:d^{3} p_{m},\\
A^{-1}& = \int \frac{1}{E_{l}}d^{3}p_{l} \int \frac{1}{E_{N'}}d^{3}p_{N'},
\label{13}\end{split}\end{align}
in the case that the final nucleon is free,
where
\[
\sigma_0(\tau, \tau')=
|g_\Delta^N \times\bra{\Delta,l}g_\tau^{\tau'}\ket{N, \nu}|^2.
\]
When the final nucleon is bound, antisymmetrization reduces
its phase-space and
\begin{eqnarray}
A^{-1} = \int \frac{1}{E_{l}}d^{3}p_{l}\: \Omega_{N'},
\end{eqnarray}
where $\Omega_{N'}$ corresponds to the final nucleon phase-space.

In CRISP model, $\Omega_{N'}$ is calculated by considering
the Pauli blocking mechanism.
So, we can normalize by setting
\begin{eqnarray}
A^{-1} = \int \frac{1}{E_{l}}d^{3}p_{l}.
\label{15}\end{eqnarray}
Finally, in this description, the only unknown quantity
is $\sigma_0 (\tau, \tau')$,
which is independent of the particle momenta. We understood
that this treatment of $\Delta$-nucleon interaction is rough and obeys
to simplicity in the present KM.
A consistent formalism of the $\Delta$-nucleon interaction
from the effective Lagrangian theory
was performed by Mariano {\it et al.} ~\cite{Barb13,Mari11}
within the CIM.

\begin{figure*}[!ht]
\centering
 \subfigure[$\mathbf {\boldsymbol{\nu_{\mu}+n\rightarrow \mu^{-}+p}}$]
{\includegraphics[scale=0.35]{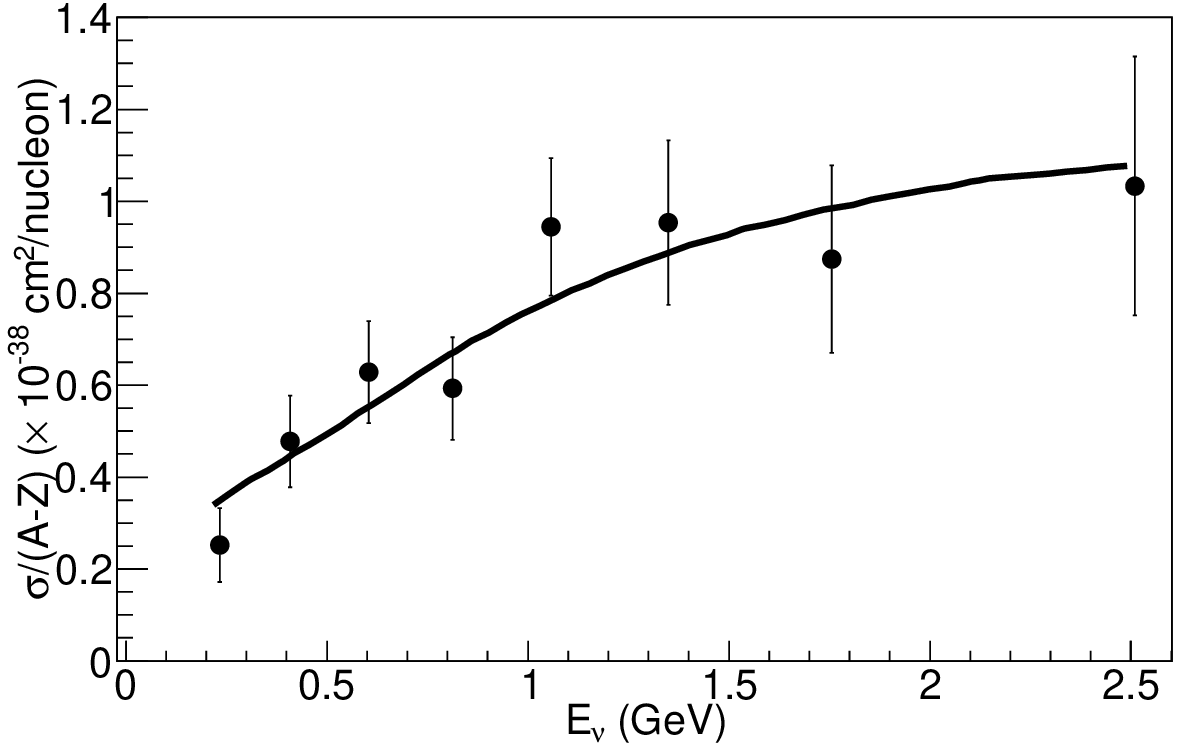}}
 \subfigure[$\mathbf {\boldsymbol{\nu_{\mu}+p\rightarrow \mu^{-}+\pi^{+}+p}}$]
{\includegraphics[scale=0.35]{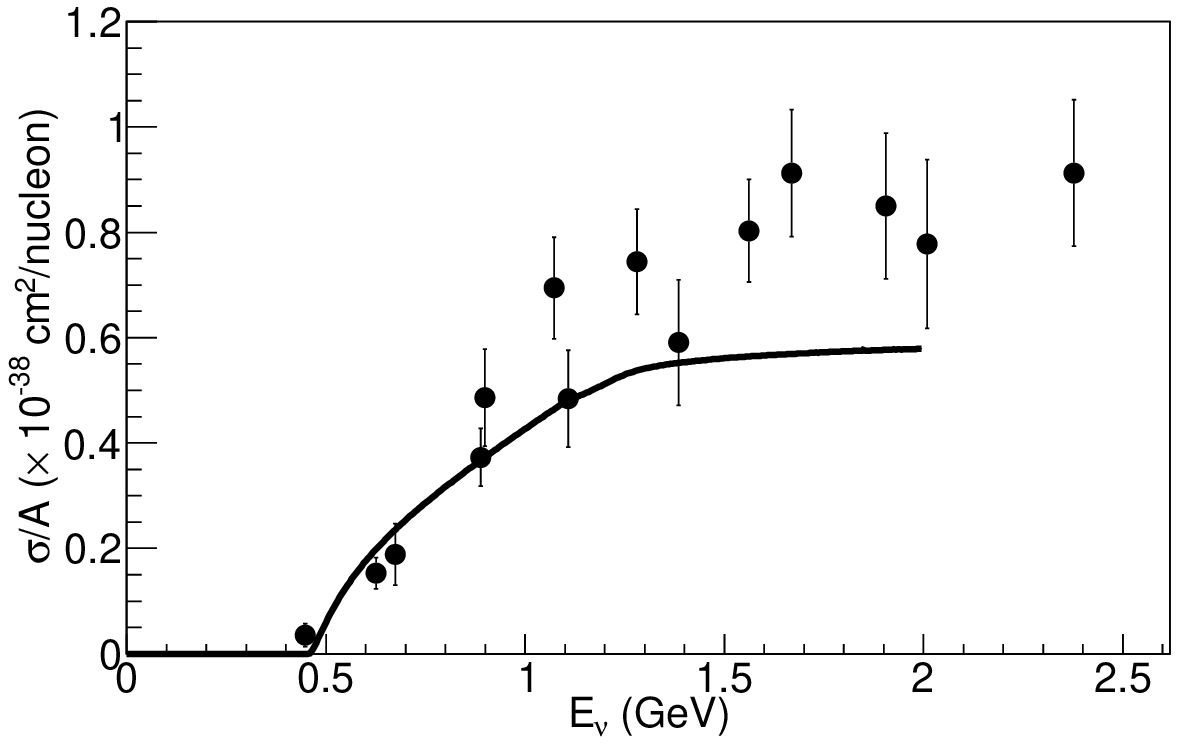}\label{figPrimInte3_7_a}}
 \subfigure[$\mathbf {\boldsymbol{\nu_{\mu}+n\rightarrow \mu^{-}+\pi^{+}+n}}$]
{\includegraphics[scale=0.35]{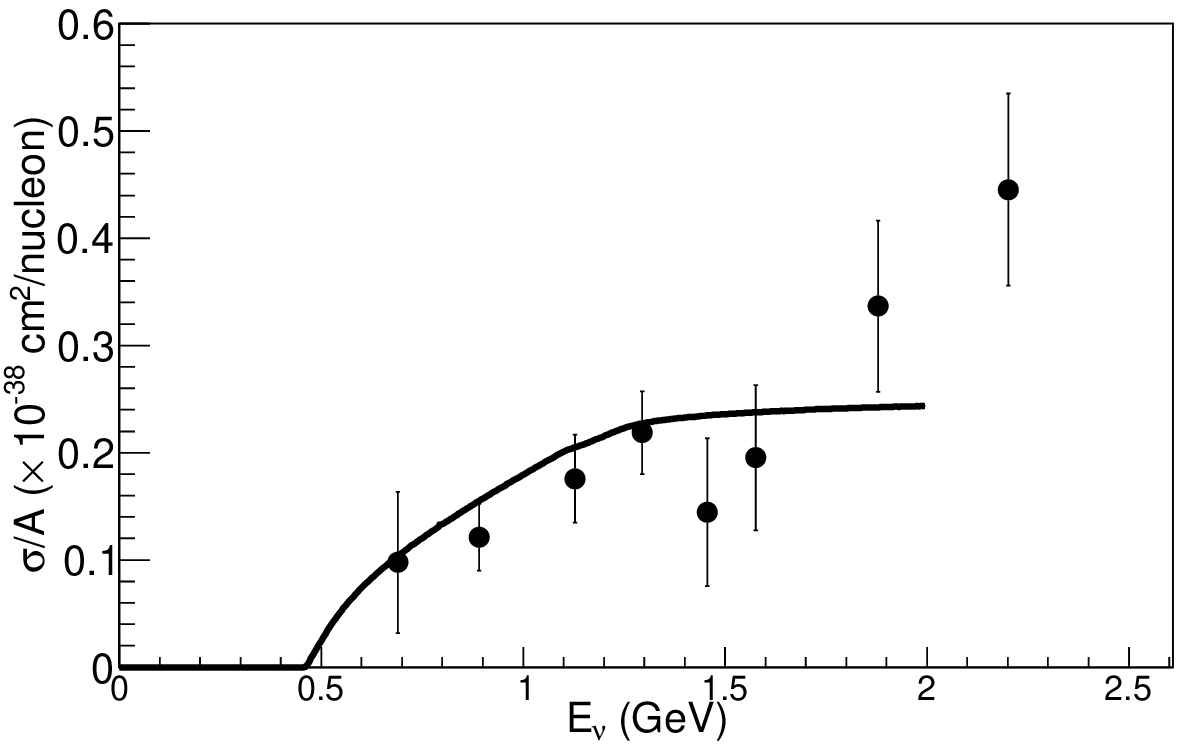}\label{figPrimInte3_7_b}}
\subfigure[$\mathbf {\boldsymbol{\nu_{\mu}+n\rightarrow \mu^{-}+\pi^{0}+p}}$]
{\includegraphics[scale=0.35]{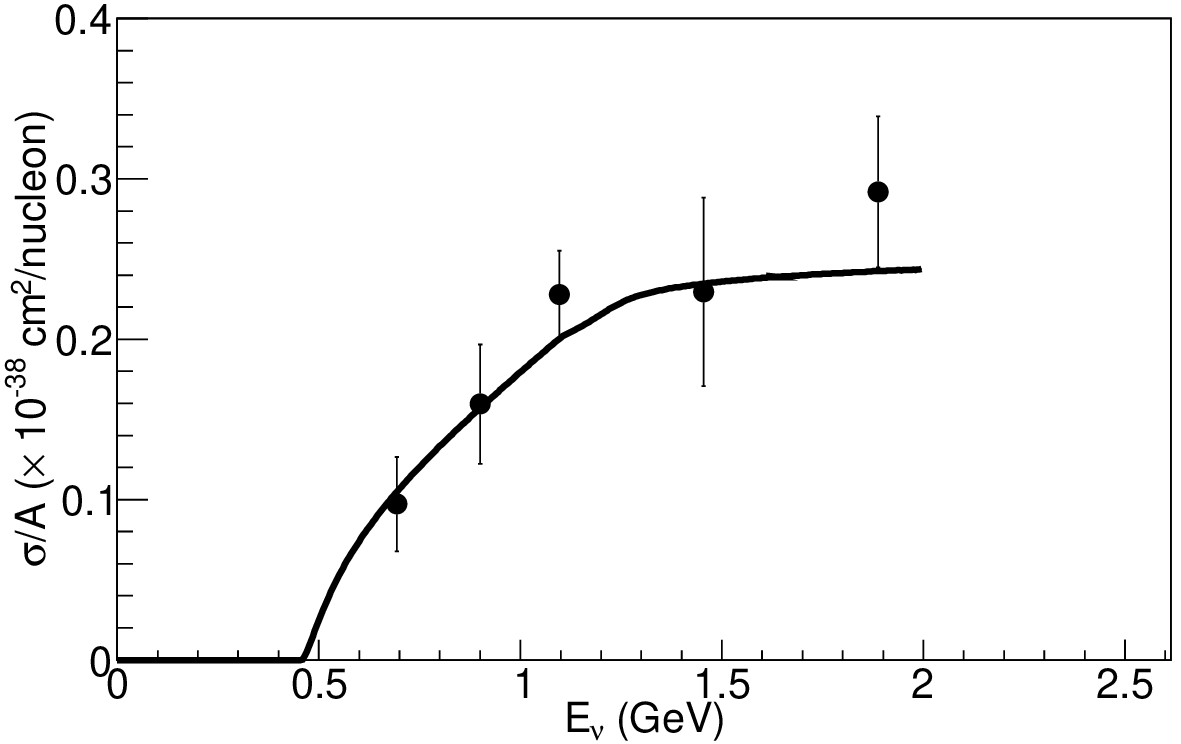}}
\caption{Best  fitted results of KM for
the $\nu_\mu$-deuterium cross
section compared with
experimental data from Refs.~\cite{BNL86,ANL79}.}
\label{figBestFit}
\end{figure*}

\subsection{On fake events}
Ones of the most relevant nuclear effects are the fake events,
{\it i.e}, states at the end of the nuclear reaction initiated
by the neutrino that are
different from the states formed by the neutrino-nucleon
primary interaction,
and if they are detected could be confused with another event.
This effect is a consequence of the final state interaction
in the nucleus.
Moreover, could to come,
for instance, from the nucleon-nucleon interaction where
is exchanged
the energy of protons and neutrons
in the binary collisions of the intranuclear
cascade or, in the case where the expected nucleon in the
final state remains bound to the nucleus.
The output result is that the original state produced
is counted in the primary interaction
as a different state.
We will refer to this effect as crossed channels or fake events, as
they are usually called in the literature.

\section{Results and Analysis \label{ResAna}}
\subsection{Free parameters adjustment }
In the KM the only free parameters are $A_n ~(n= 0, 1 , 2, 3, 4)$
for the CCqe channel and $\sigma_o(\tau,\tau')$ for the CCres channel.
In order to determine these parameters,
neutrino-deuterium cross section on deuterium measured at Brokkhaven
National Laboratory (BNL)~\cite{BNL86} and Argone National
Laboratory (ANL)\cite{ANL79} were used.
This information is
the same experimental data employed
previously by O. Lalakulich and U. Mosel in their studies
on pion production in the MiniBooNE
experiment ~\cite{Lalakulich2013}.
Here, we disregard the small nuclear effects present in
the interaction with deuterium and consider the cross section
as representative of the neutrino-nucleon process.
In Figure~\ref{figBestFit} we present the best-fitted result for
our model to the available
experimental data. In Table~\ref{canais} the corresponding values
for all the parameters are displayed. One observes a nice fit of
our calculation to the experimental data. In
the case of the channel in Figure~\ref{figPrimInte3_7_a}
(channel B of Table \ref{canais})
one can notice that the calculation slightly underestimates data above
$E_\nu \sim 1.5$ GeV.
This result can be attributed to the lack of resonances
heavier than $\Delta(1232)$ in the present version of our model.

With the inclusion of the KM described in the last section
into the CRISP model, we can evaluate the nuclear effects on the
neutrino-nucleus interaction and calculate
the inclusive neutrino-nucleus cross section
up to $E_\nu \sim 1.5$ GeV.
	
\subsection{Reaction cross section}
The neutrino-nucleus cross sections are
determined employing
the CRISP code by
calculating the frequency of
appearance of a
previously obtained channel from
a number $N_0$  of total events.
So then, the cross section reads
\begin{eqnarray}
\sigma_{ev}=\sigma_g \frac{N_{ev}}{N_0},
\label{16}\end{eqnarray}
where $N_{ev}$ is the number of events that ended
 within the concrete
channel under analysis,
and $\sigma_g= \pi r_0^2 A^{2/3},$
is the nucleus geometric cross section, with $A$ being
the mass number and $r_0= 1.2$ fm.

 We consider as an event, to anything
of the final configuration listed in Table~\ref{canais}.
The calculations are performed for the two kinds of events,
namely true-type and like-type
events.  A true-type event is when the final configuration
is exactly as those
listed above, while  a like-type event is when the configuration exists
amid other particles.

\begin{figure*}
\centering
\subfigure[]{\includegraphics[scale=0.4]{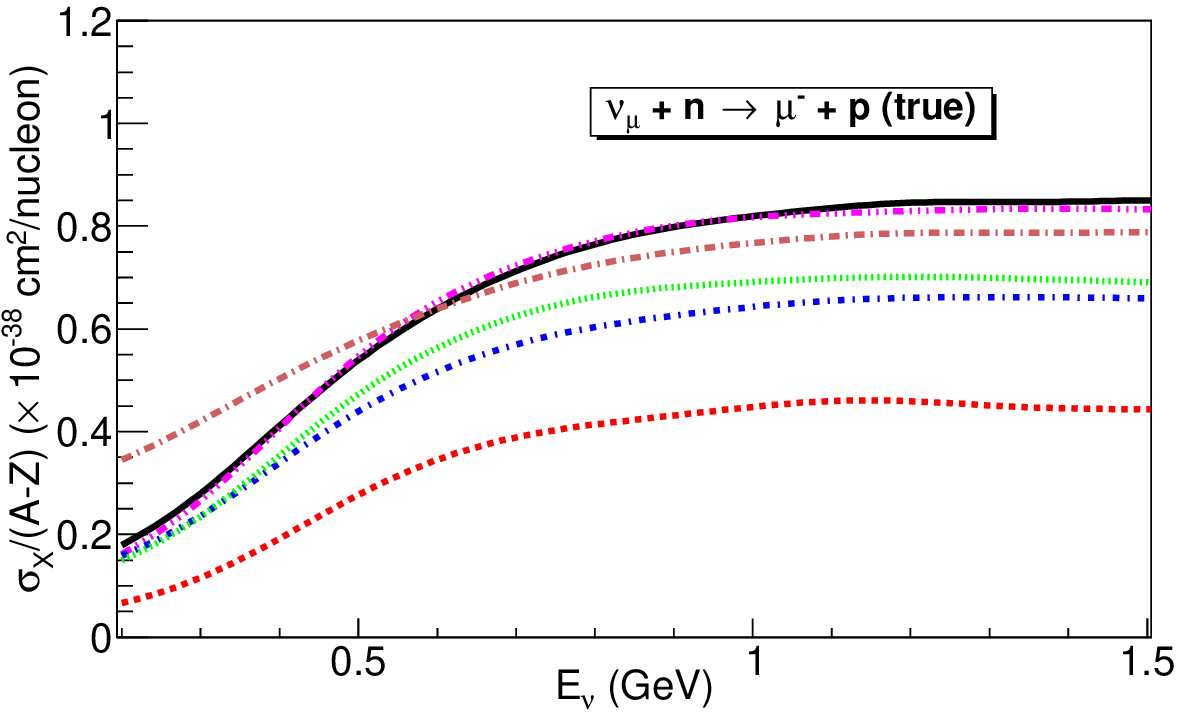}}
\subfigure[]{\includegraphics[scale=0.4]{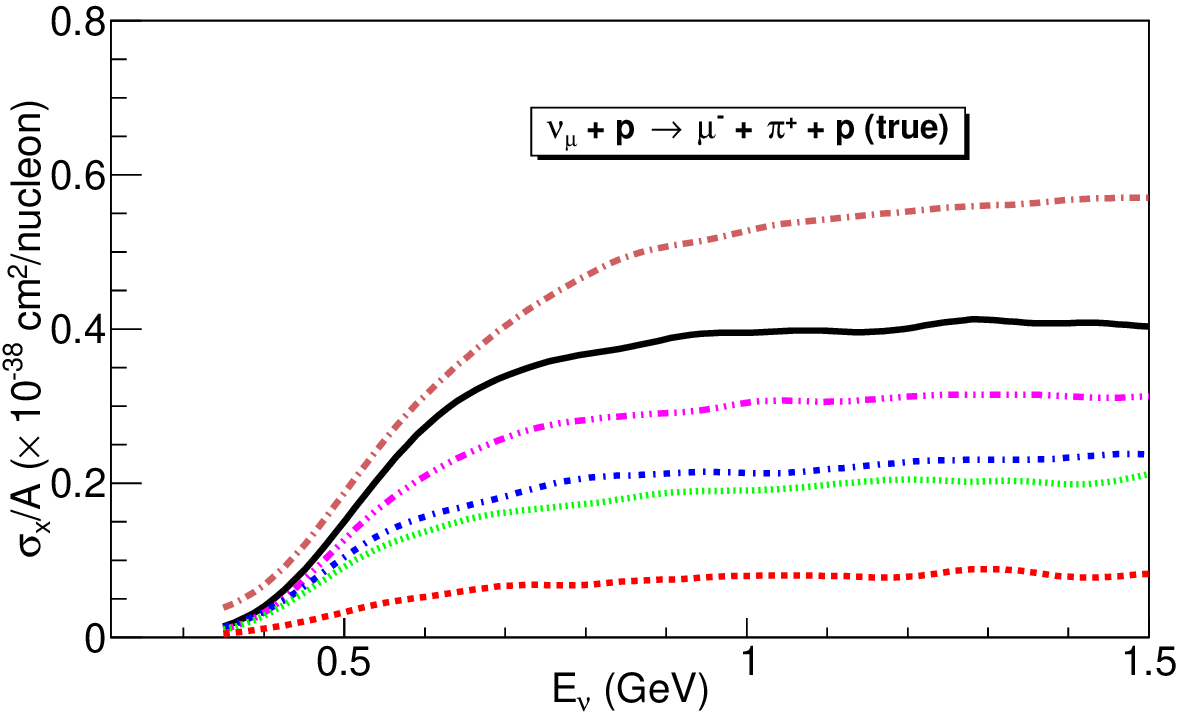}}
\subfigure[]{\includegraphics[scale=0.4]{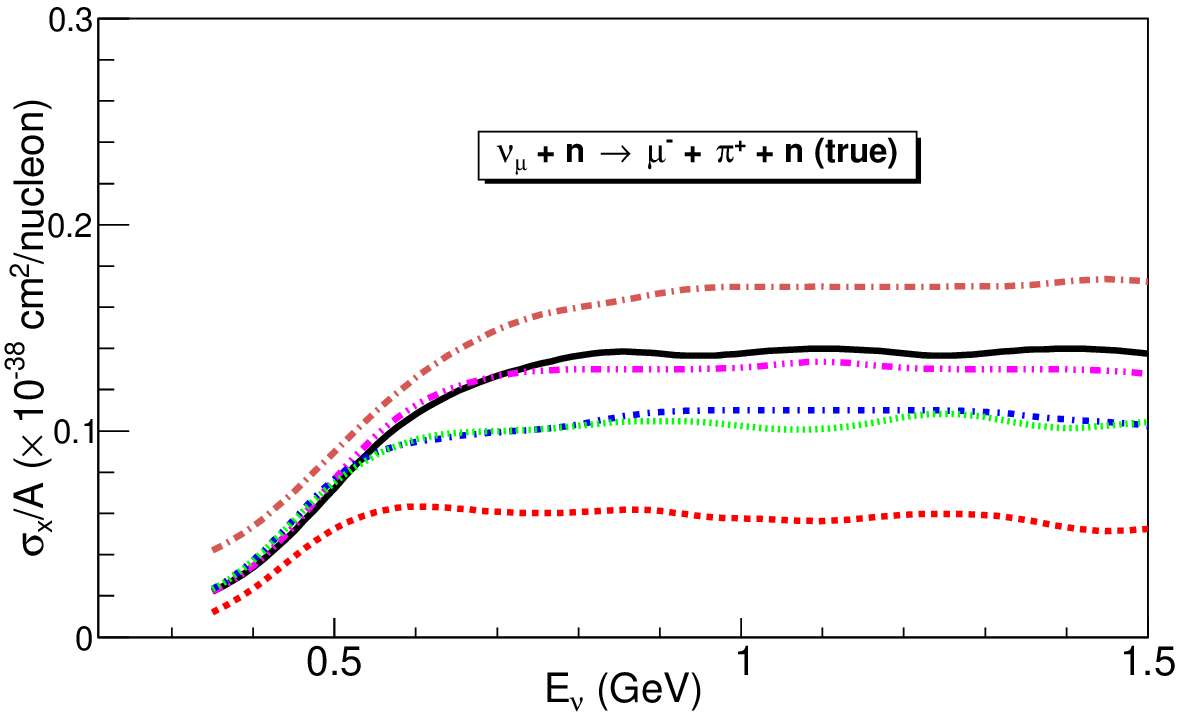}}
\subfigure[]{\includegraphics[scale=0.4]{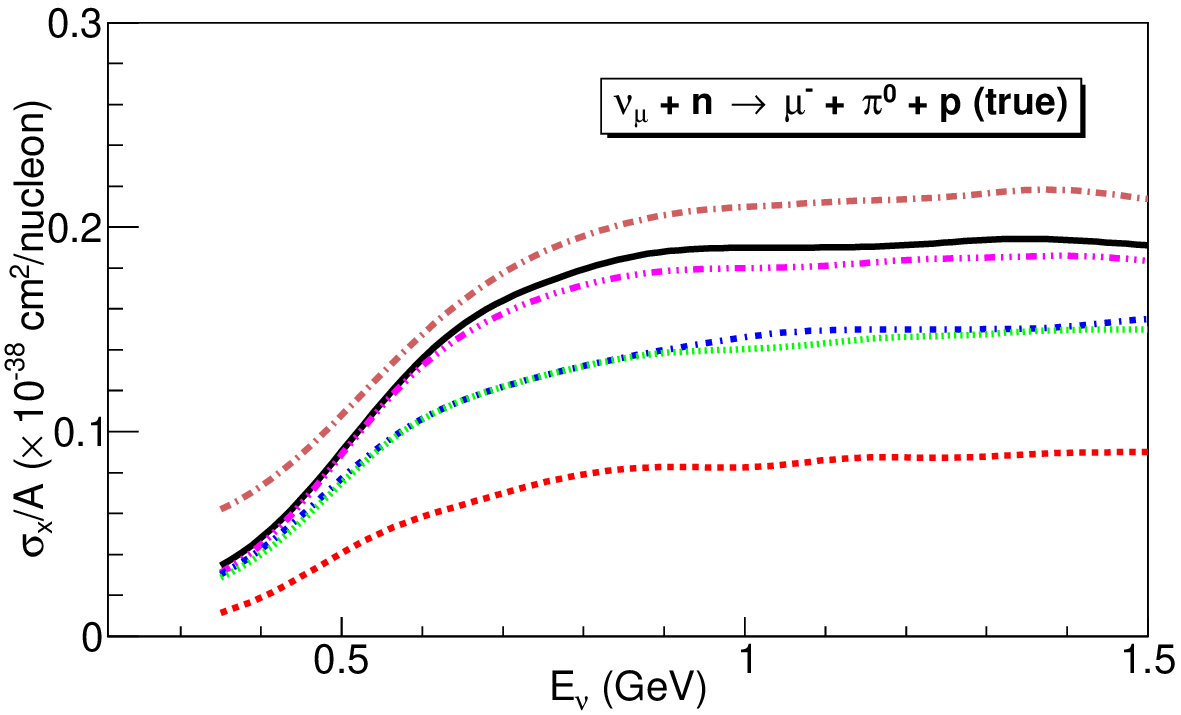}}
\caption{(Color online) $\nu_\mu$-A cross sections for all true-type channels
studied as function of neutrino energy for all nuclei:
(long dash dot brown line) $^{12}$C, (solid black line)$^{16}$O,
(dash dot rouge line) $^{27}$Al, (dash dot blue line) $^{40}$Ar,
(dot green line) $^{56}$Fe, and (dot red line) $^{208}$Pb.}
\label{figCS}
\end{figure*}

The total cross sections for each channel are shown in
Figure~\ref{figCS} for all nuclei studied in the present work.

\begin{figure}[!ht]
\centering
\includegraphics[scale=0.42]{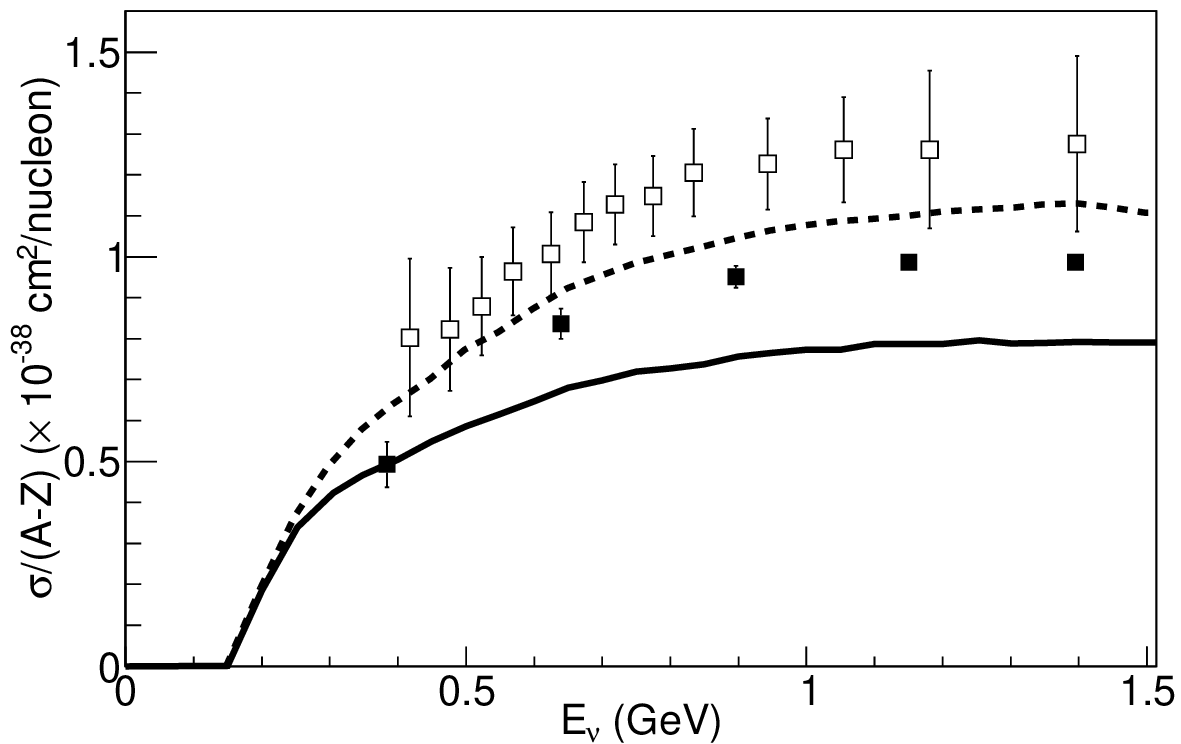}
   \caption{$\nu_\mu$-$ ^{12}$C cross section within  KM model according
true-type (solid line) and like-type (dashed line) events
for $\mathbf {\boldsymbol{\nu_{\mu}+n\rightarrow \mu^{-}+p}}$ channel.
The experimental data for true-type (filled squares)~\cite{Eberly15} ,
and like-type (hole squares)~\cite{MiniBooNE11}
are shown for comparison.}
\label{fig12C_CS} 
\end{figure}

In Figure~\ref{fig12C_CS} 
the calculated CCqe cross section are compared with
the available experimental data on $^{12}$C. We observe an overall
agreement between calculation and experimental data in
the like-type events, while on the
true-type events the theoretical calculations underestimate the data.
The diminishing in the nuclear cross
section in the last case is mainly attributable to events
where the proton in the final state remains bond in the nucleus.

In the CRISP model, the nucleus is described  as
a Fermi gas in a square-well potential, which is a fair description
for heavier nuclei, but is not adequate for $^{12}$C. Unfortunately,
there are not any experimental
data for other nuclei or channels, but one can expect that this
nuclear structure effect is less important for heavier nuclei.

\begin{figure*}[!ht]
\centering
\subfigure[]{\includegraphics[scale=0.35]{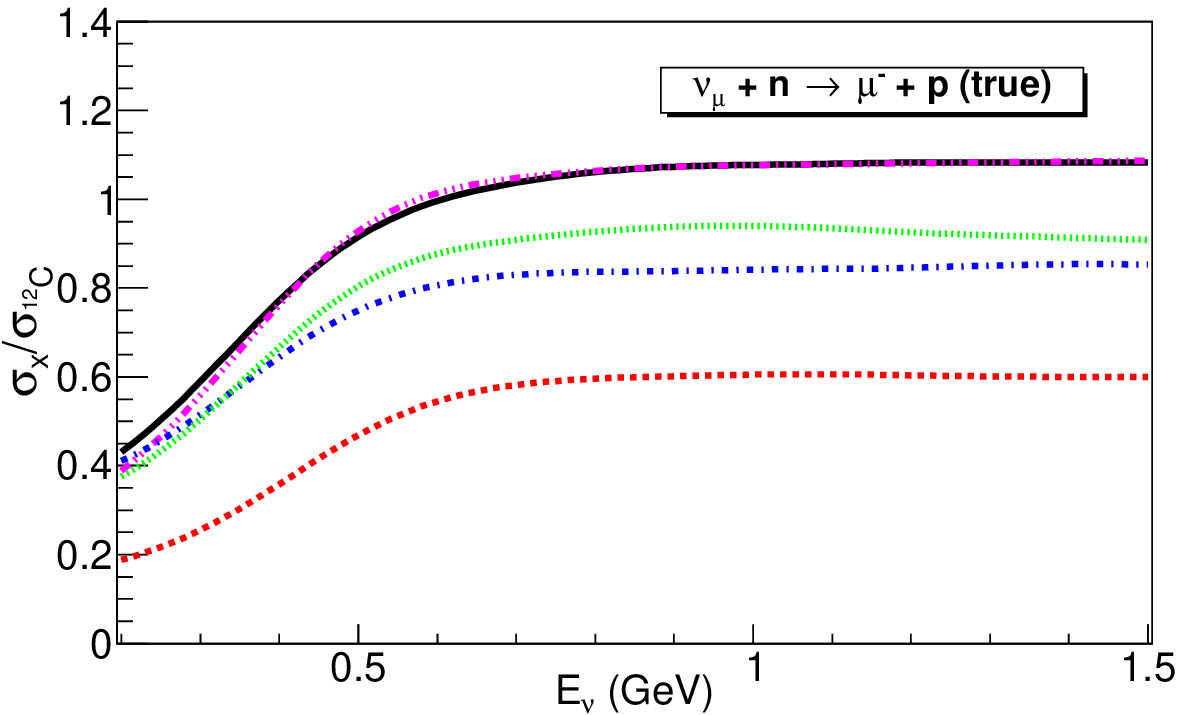}}
\subfigure[]{\includegraphics[scale=0.35]{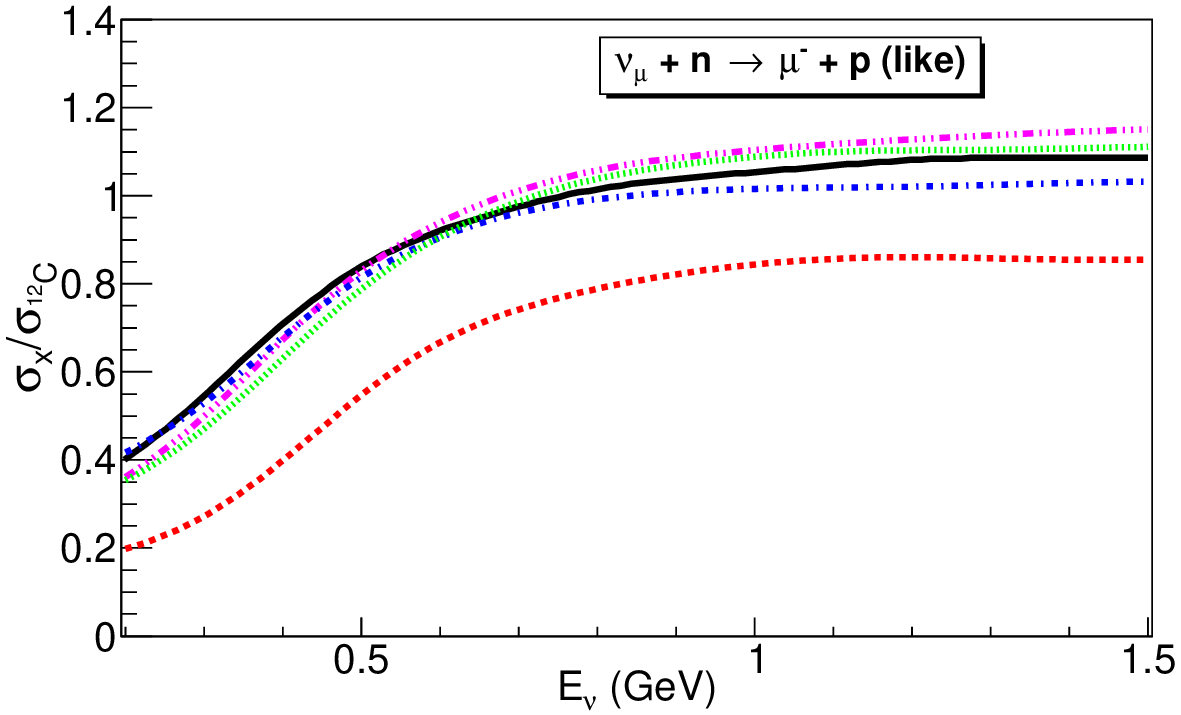}}
\subfigure[]{\includegraphics[scale=0.35]{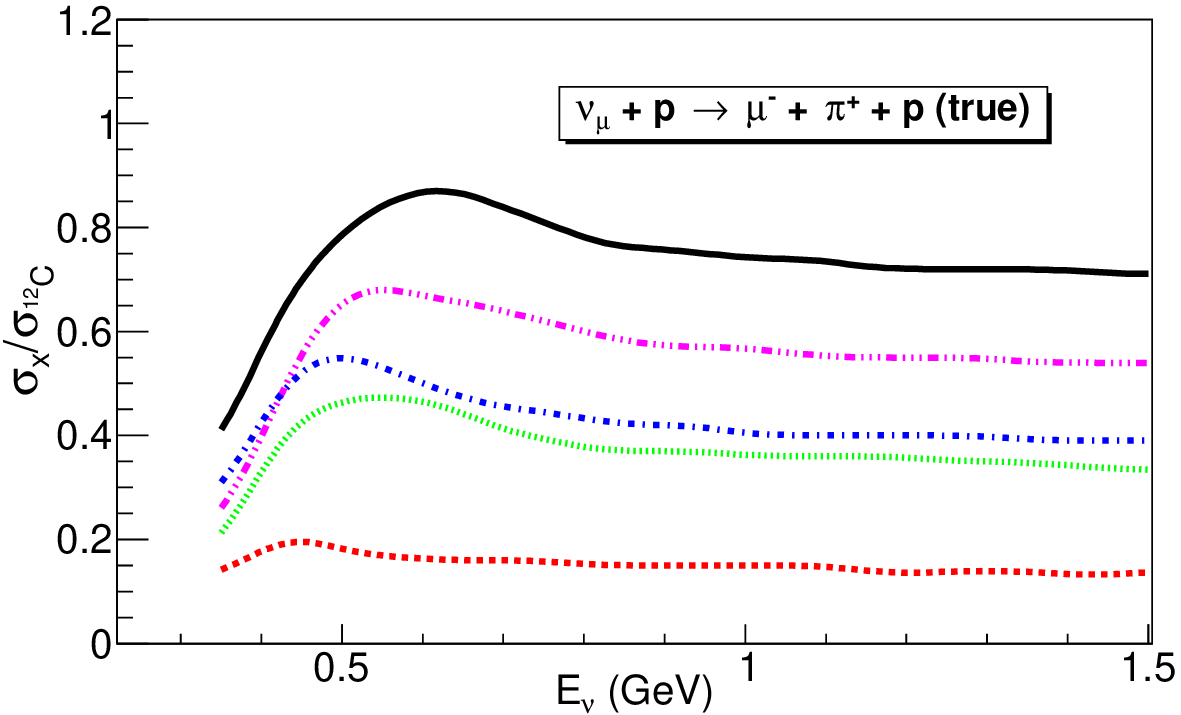}}
\subfigure[]{\includegraphics[scale=0.35]{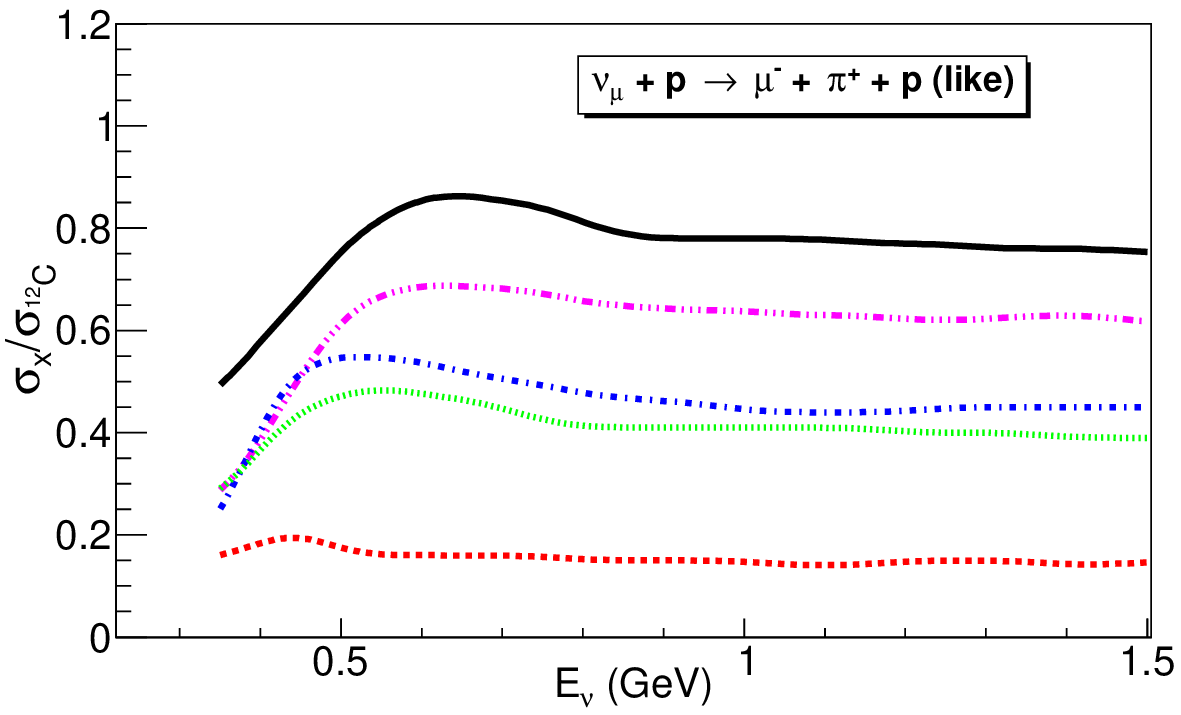}}
\subfigure[]{\includegraphics[scale=0.35]{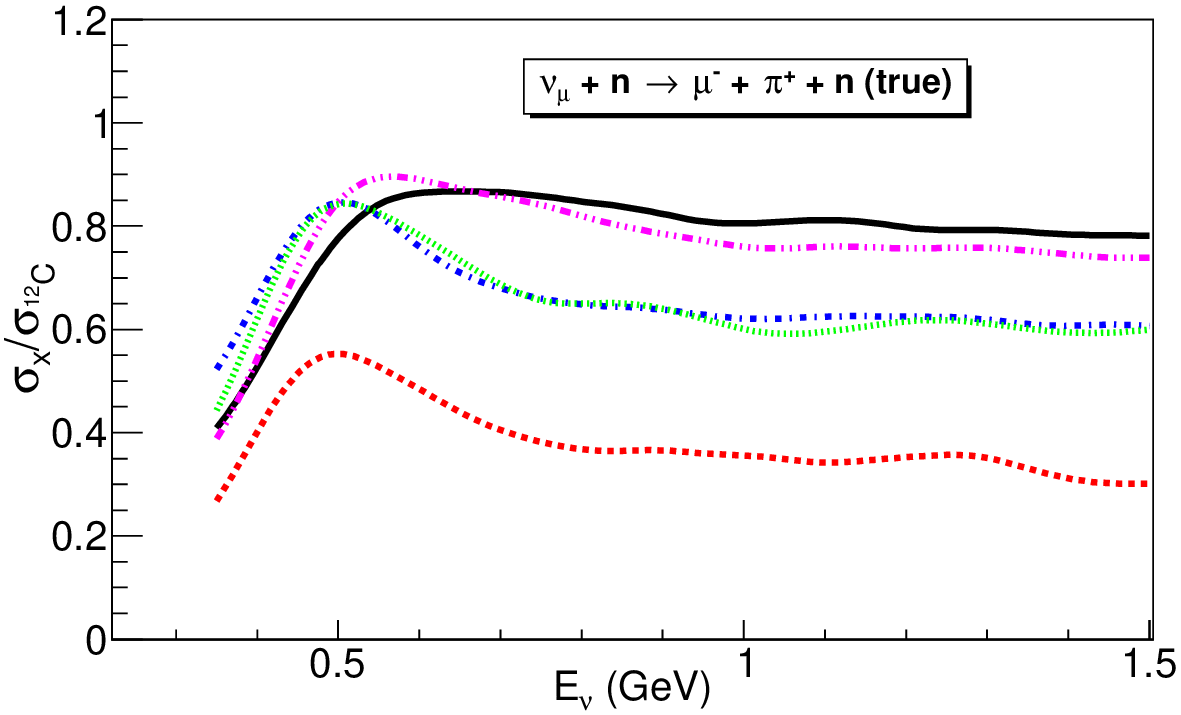}}
\subfigure[]{\includegraphics[scale=0.35]{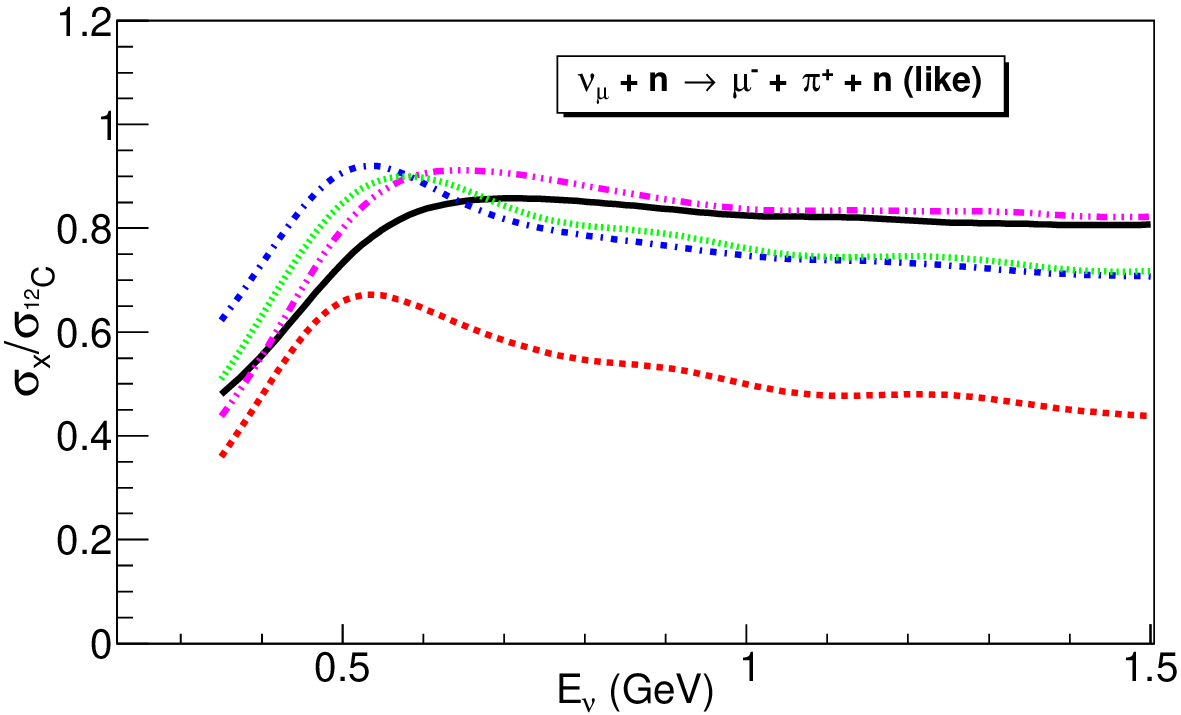}}
\subfigure[]{\includegraphics[scale=0.35]{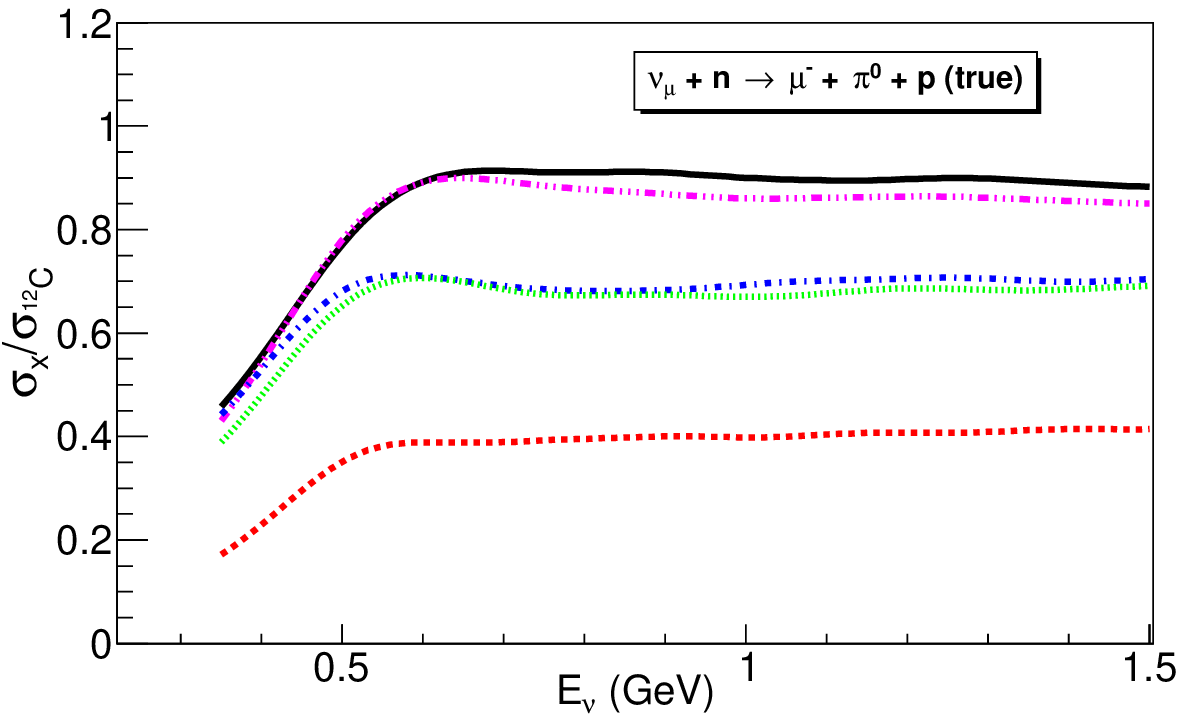}}
\subfigure[]{\includegraphics[scale=0.35]{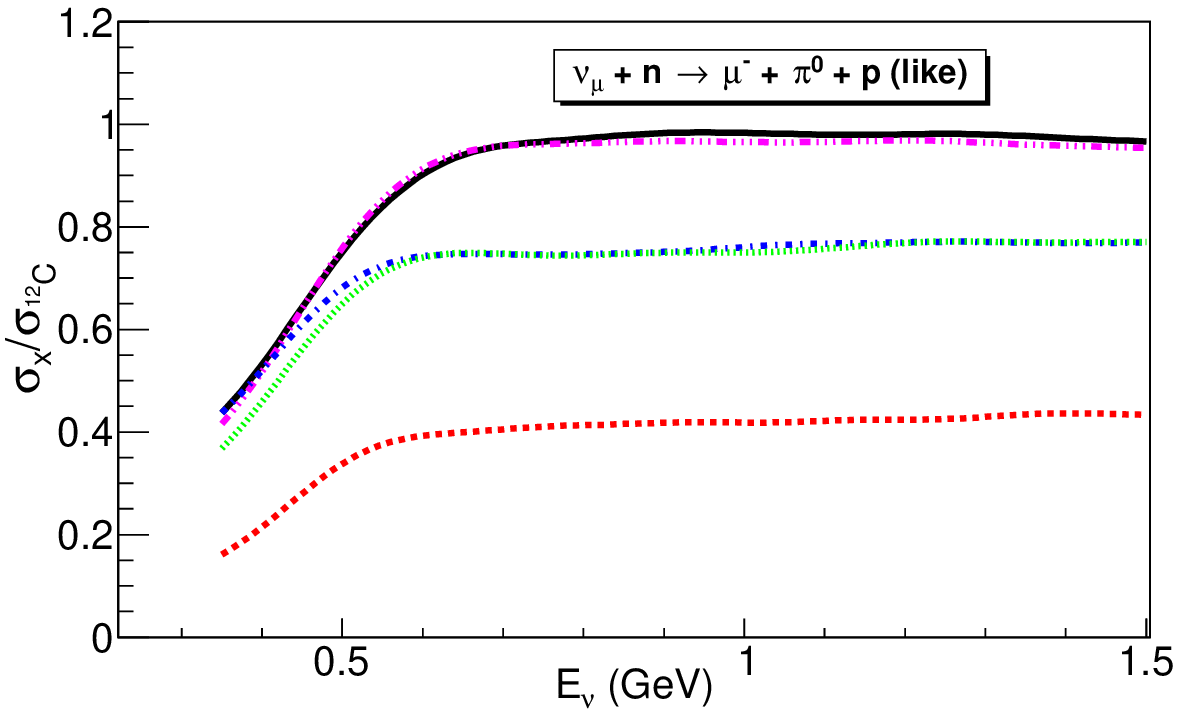}}
\caption{ (Color online)
Inclusive muon neutrino-nucleus cross section ratios
relative to $^{12}$C, ${\cal R}= {\sigma_X}/{\sigma_{12C}}$,
for different channels
as function of neutrino energy: (solid black line)$^{16}$O,
(dash dot dot dot rouge line) $^{27}$Al, (dot green line) $^{56}$Fe,
(dash dot blue line) $^{40}$Ar, (dot red line) $^{208}$Pb.
Left panel: true-type
and Right panel: like-type events.
}
\label{fig_CSrelative}
\end{figure*}

One can notice in Figure~\ref{figCS} that there is an overall
decreasing in the cross
sections when the nuclear mass increases.
Although the nuclear level structure here considered, corresponding
to a square-well potential,  it
has some noticeable effects on the relative
cross sections.
To better visualizes this effect, in Figure~\ref{fig_CSrelative}
we present the calculated cross section ($\sigma_X$)
normalized to the $^{12}$C cross section ($\sigma_{12C}$)
for six different nuclei: $X=\{^{16}{\rm O}, ^{27}{\rm Al},
^{40}{\rm Ar}, ^{56}{\rm Fe}, ^{208}{\rm Pb}\}$,
for both true-type and like-type events. Two general aspects
are
observed: (i) there is a  fast increase
in ${\cal R}=\sigma_X/\sigma_{12C}$
for all nuclei at low neutrino energy
and, (ii) there is
an  explicit
dependence on the nuclear  mass.

The diminishing
in the cross section
is due to nuclear effects, since it depends on
the nuclear mass. In fact, as the nucleon produced in the final
state after the
neutrino-nucleon interaction propagates inside the nuclear matter,
it can transfer
its energy to other nucleons, so in many cases, the particles
emitted from the
nucleus are not exactly those formed in the neutrino-nucleon interaction.

Since the average length of the distance traveled by a nucleon
increases with the nuclear mass, also the probability of crossed channels
increases.
This effect reduces, therefore, the observed cross section, as can be
noticed in Figure~\ref{fig_CSrelative}.

The rise in the ratio
of low energies, ${\cal R}$, is related to
the escape of the nucleon produced in the primary interaction
of the nucleus.
At low energy, most of the neutrino energy goes to the muon production,
leaving the nucleon with low energy, so it
can not overcome the nuclear barrier.
The result is that  one has a crossed channel event. Over of the region
of fast increase of  ${\cal R}$, appears a plateau that  it remains
approximately constant for all nuclei and channels. In this region,
the produced nucleon
has enough energy to escape from the nucleus. However, in its way out the
nucleon will interact with others nucleons and eventually,
a  charge-exchange collision will produce a crossed channel.
Then, as larger is the nucleus as
higher is the probability of crossed channel events,
and for the CCqe case, we have checked that this probability
is roughly proportional to $A^{1/3}$.

The same reasoning applies qualitatively to the resonant channel.
However,
in these cases, the reaction mechanism is more
complicated
because the resonance
propagates inside the nucleus exchanging energy  with other nucleons,
which will produce other effects that superpose to the ones described above.
For example in Ref.~\cite{Cuyck2016}, the authors have
presented detailed calculations performed in $^{12}$C showing that
the two-particle two-hole
is the mainly contribution of multinucleon
excitations. Many of these effects are related by
the influence of short range correlations (SRCs) on the one-nucleon (1N)
and two-nucleon (2N) knockout channels, and to two-body currents
arising from meson-exchange currents.
For the channel of the like-type, the results for ${\cal R}$
are very similar to the true-type.

Another real nuclear effect on the neutrino-nucleus interaction
can be observed in the calculated cross sections for nuclei,
as shown in
Figure~\ref{fig12C_CS} 
, as compared to the interaction on nucleon, as shown in
Figure~\ref{figCS}.  
In fact,  one can observe that the interaction threshold is
around 0.45 GeV in the
nucleon case, while the threshold is below that energy for
nuclear interactions.
This subthreshold interaction is due to the Fermi motion
of bound nucleons,
and it is a natural consequence of our calculations using the CRISP model.
In fact, these kinds of phenomena can also
be observed
for other processes~\cite{Israel2014}.
Goldhaber and Shrock \cite{Goldhaber2001} relating
by first time the possible
subthreshold reactions involving nuclear fission such as:
(i)  photo-fission with pion-production and,
(ii) charged-current neutrino-nucleus
reactions that lead to fission and/or to the
formation of a Coulomb bound state of a  $\mu^-$
with the nucleus of a fission fragment,
that is a very similar to the reactions studied in this work.

\begin{figure*}[!ht]
 \centering
 \subfigure[]{\includegraphics[scale=0.42]{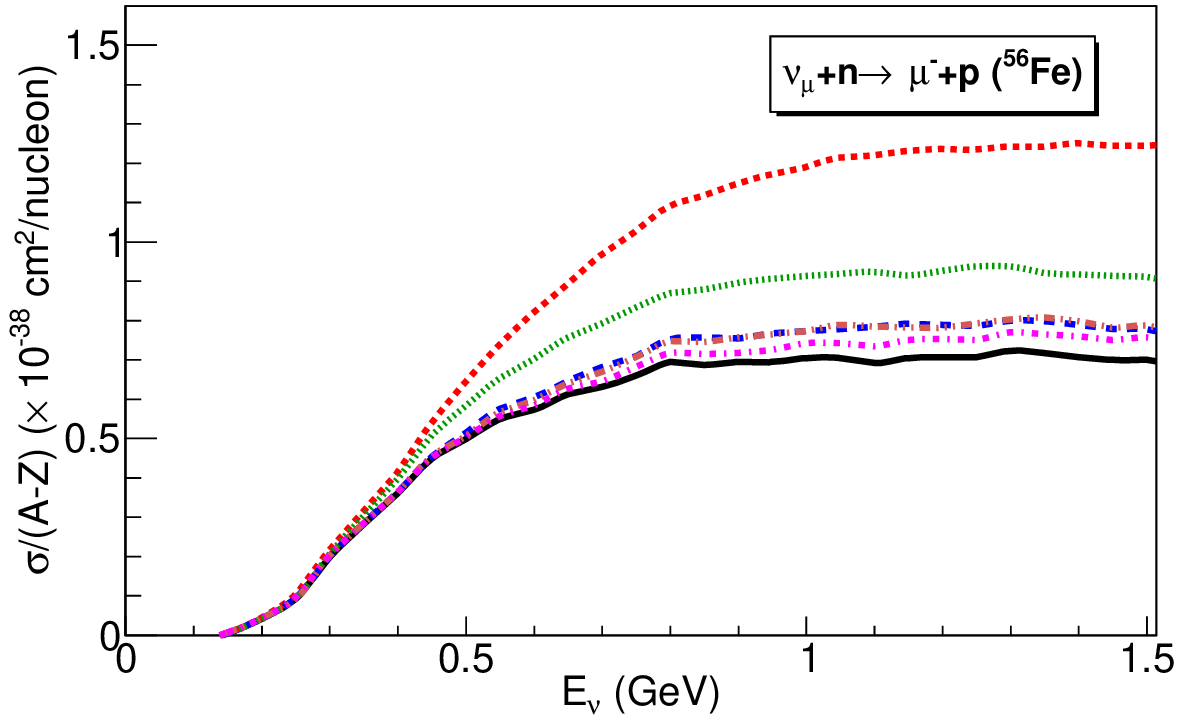}}
 \subfigure[]{\includegraphics[scale=0.42]{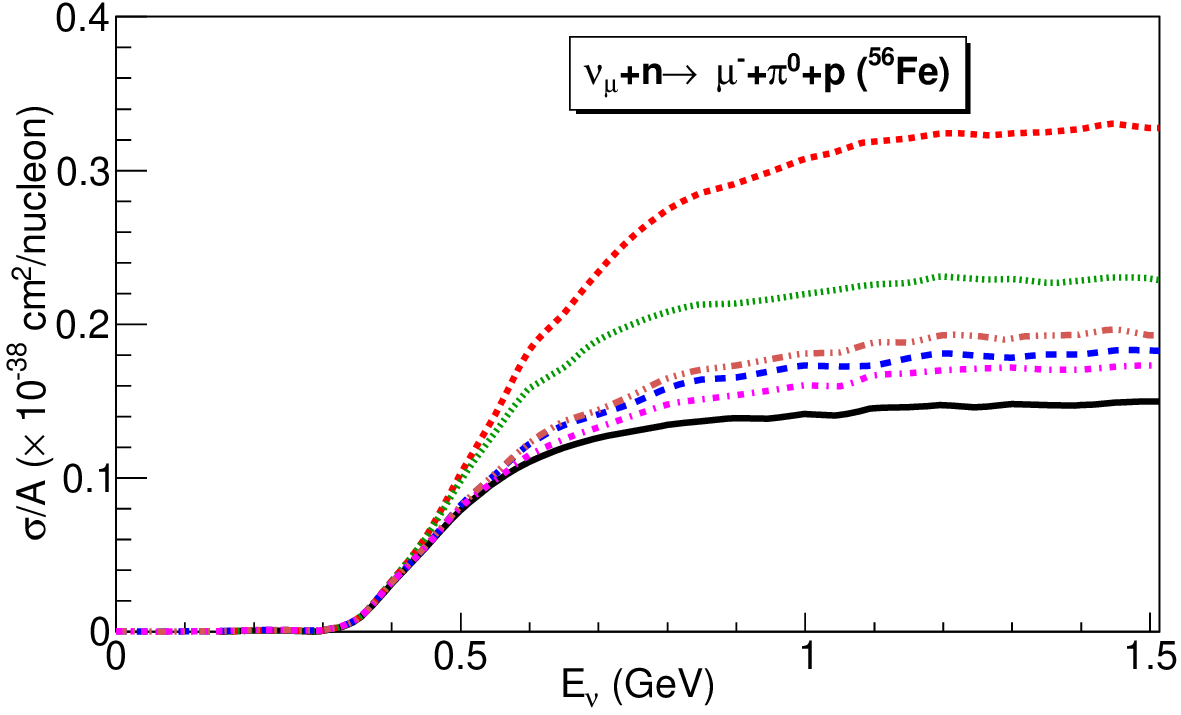}}
 \subfigure[]{\includegraphics[scale=0.42]{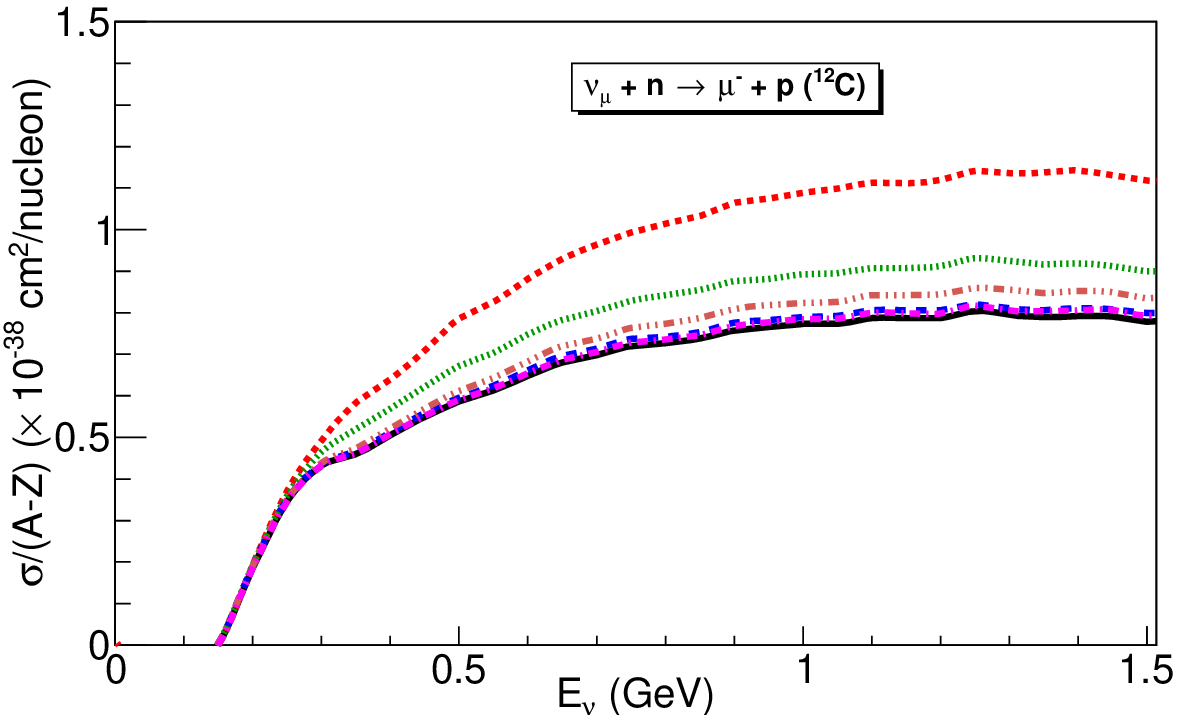}}
 \subfigure[]{\includegraphics[scale=0.42]{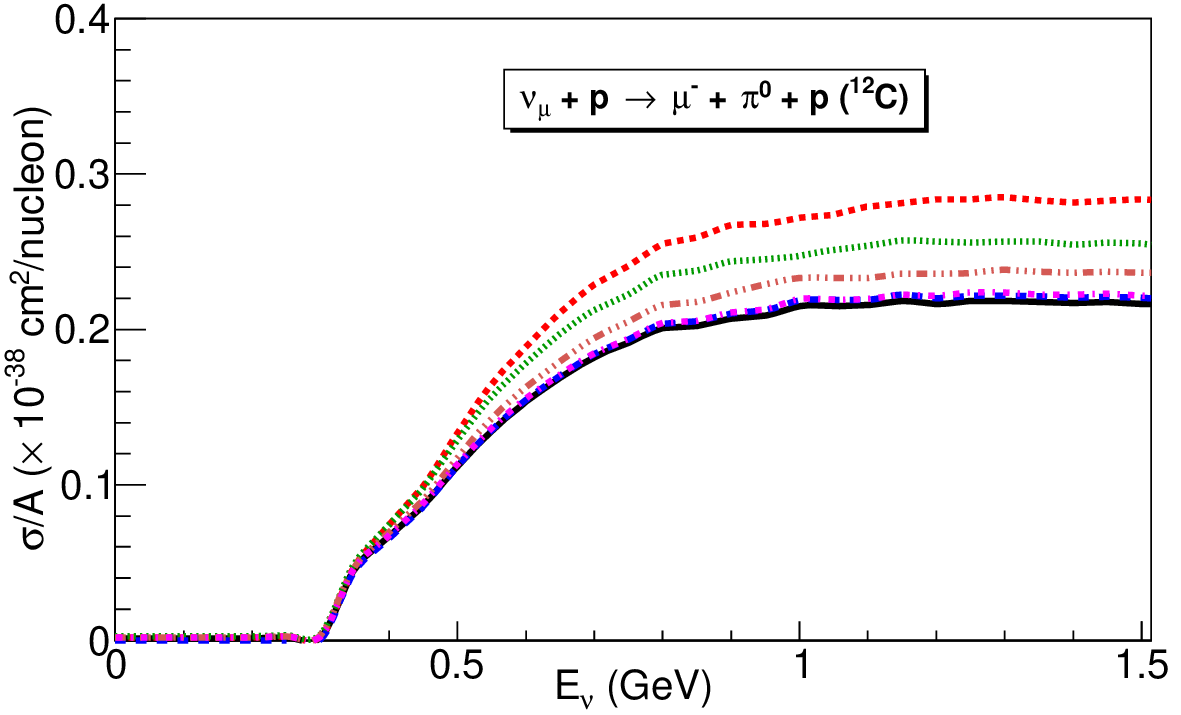}}
\caption{(Color online) Partial contributions from some final state
configuration to the like-type cross section for $^{56}${Fe}
and $^{12}$C in the CCqe channel (left panel)
and the CCres channel (right panel).
We show the contributions: added like-type (dash red line),
1p+1n (dot green line), 2p+1n (dash dot brown line),
1p+2n (long dash blue line),
2p+2n (dash dot pink line), and finally the
true-type (solid black line).}
\label{fig_partialContr} 
\end{figure*}

In Figure~\ref{fig_CSrelative} we analyze the like-type events
as compared to the true-type ones.
The like-type cross
sections are always higher since these
counts the true-type events and also more complex configurations.
With the  CRISP model, it is possible to disentangle the various
contribution to the like-type cross section, as shown in the
Figure~\ref{fig_partialContr}
for some of the more simple configurations.
In such figure, we showed
the partial contributions from some final state configuration to the
like-type cross section for $^{56}$Fe and $^{12}$C in the CCqe channel
and
the CCres channel like-type. The addition of
like-type,  1p+1n,  2p+1n,
1p+2n,  2p+2n in (dashed lines), and the true-type  in (solid line)
are also
shown for comparison. In both nuclei in consideration, the sequence
of contributions is similar:
main contribution from 1p+1n,
in second place 2p+1n plus 1p+2n, and finally
the 2p+2n, more closed to the true-type in the last reactions.
For the CCqe reactions the true-type and like-type are broadening
in $\sim 0.3$ GeV, whereas for CCres this
"threshold" is in $\sim 0.4$ GeV  due the nuclear
delta liberty grade.

As one allows more and more complex configurations,
the cross  section rises from the true-type cross section
to the like-type cross section,
where  are considered
all possible configurations.
It is also possible
to observe that increasing the neutrino energy, the complexity
of like-type events
increases, while at low energy true-type and like-type almost coincides.
Comparing the CCqe and the resonant channels, we note that the several
trends and relative contribution of a different channel are similar.

In general, one can see that
real nuclear effects are of great importance to
understand the neutrino-nucleus interaction. These
results
are relevant since some cross section ($^{208}$Pb for instance) can be
reduced to about 20\% of the $^{12}$C one due to these effects.

The interaction with the nucleus of the particles produced in the
primary interaction is responsible for the more intense effects,
reducing the cross
section per nucleon as the nuclear mass increases. At low energy,
however, the binding energy is more important, and probably
nuclear structure will be
necessary
to completely understand the process for neutrino
energy up to $\sim 0.5$ GeV.
In this direction, a recent work within the
Continuum Random Phase Approximation (CRPA)
has calculated the $(\nu_\mu/\bar{\nu}_\mu)-^{12}$C cross section
in kinematics conditions for MiniBooNE and T2K.
The cross sections have shown to be comparable with the experimental data,
but underestimating the MiniBooNE data for backward muon
scattering angles,
where the missing strength can be associated with the contribution
from multinucleon knockout and single-pion production processes.
Among other microscopical models that can be useful in this
region is the Relativistic Quasiparticle RPA \cite{Samana2011},
that studied
the evolution of the configuration space number below 0.3 GeV.
In this energy interval,  the cross
sections  converge for sufficiently large configuration space
and final-state spin
and could be joined smoothly with the Relativistic Fermi Gas
including at least $1N$
and $2N$ knockout reaction in the same way as in Ref.~\cite{Cuyck2016}.

 \subsection{Analysis of fake events/crossed channels}

With the CRISP model used in the present analysis,
it is possible to evaluate
the amount of crossed channels in the neutrino-nucleus interaction.
This process is done by counting the number of primary
events in the
channel ``i'', $N_{p}(i)$, and the number of those events
that remains in the channel
after the intranuclear cascade is completed, $N_{f}(i)$. The number of
crossed channels events is given by
\begin{eqnarray}
N_{c}(i)=N_{p}(i)-N_{f}(i),
\label{17}\end{eqnarray}
also,
the fraction of crossed channels events is
\begin{eqnarray}
R_{c}(i)=\frac{N_{c}(i)}{N_{p}(i)}.
\label{18}\end{eqnarray}

\begin{figure*}[!ht]
\centering
\subfigure[]{\includegraphics[scale=0.35]{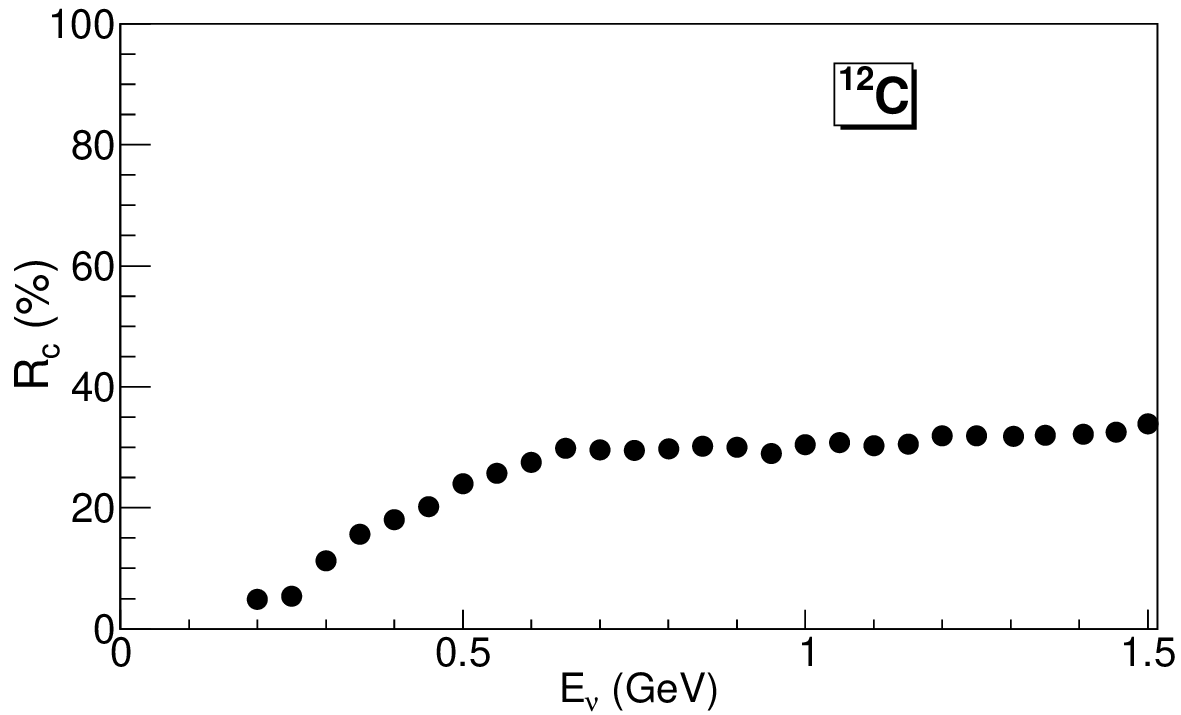}}
\subfigure[]{\includegraphics[scale=0.35]{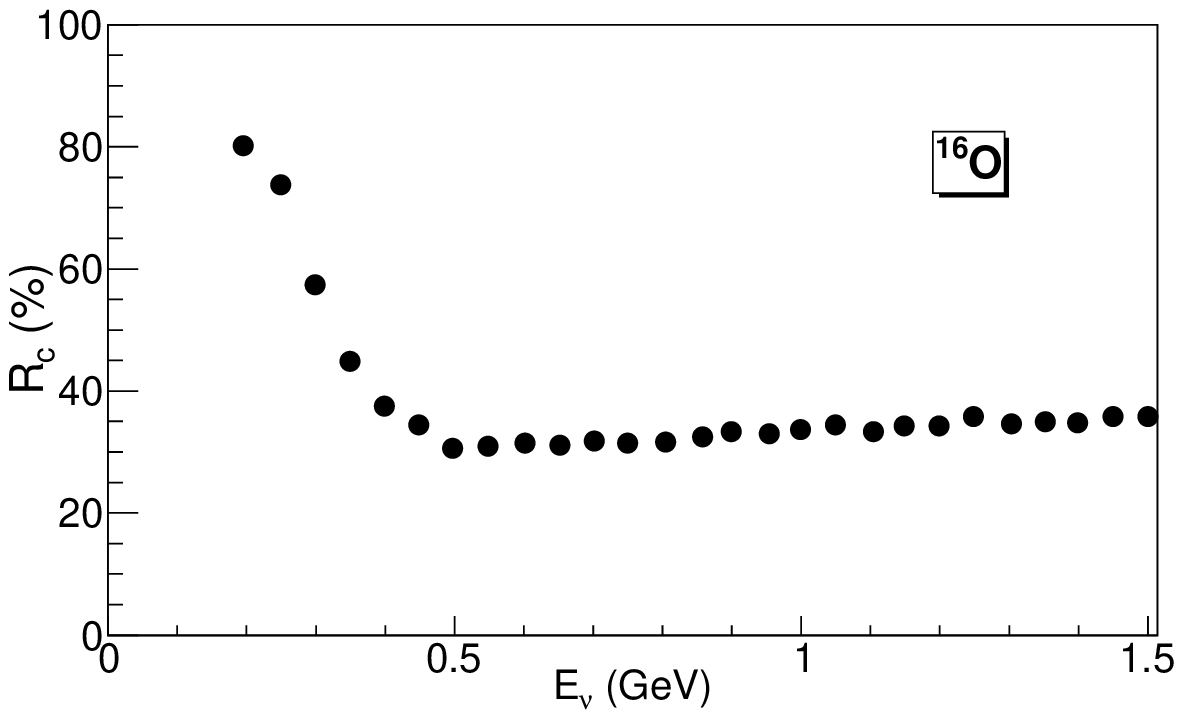}}
\subfigure[]{\includegraphics[scale=0.35]{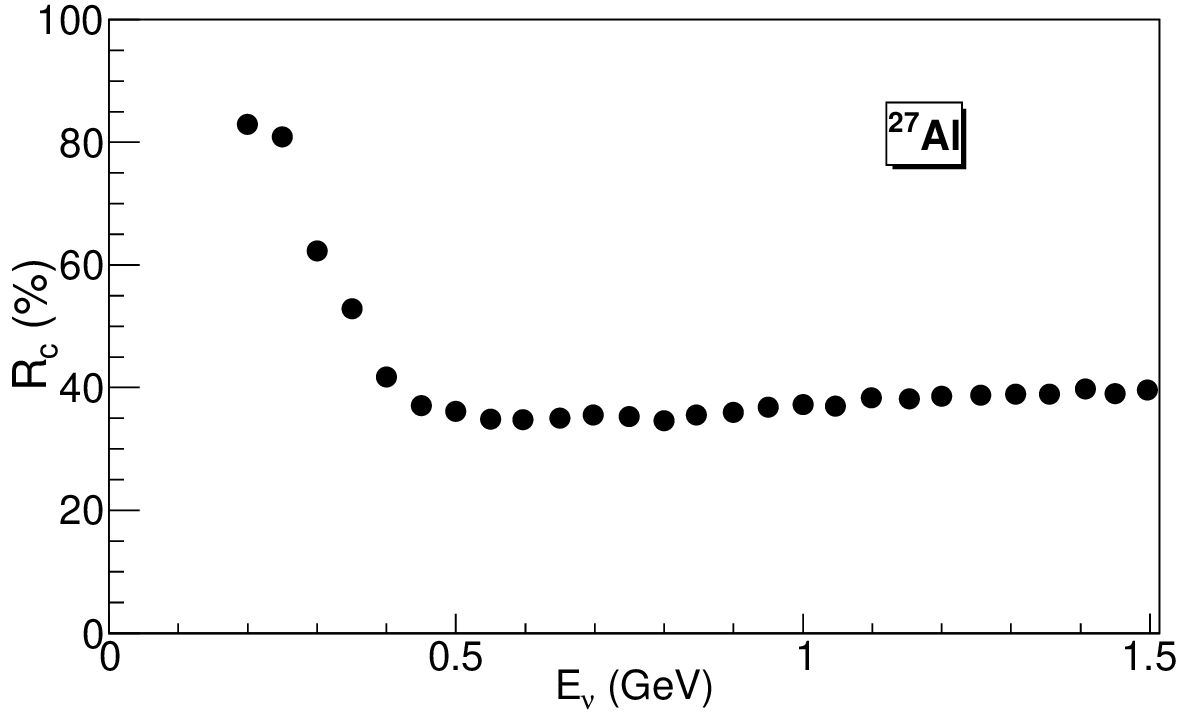}}
\subfigure[]{\includegraphics[scale=0.35]{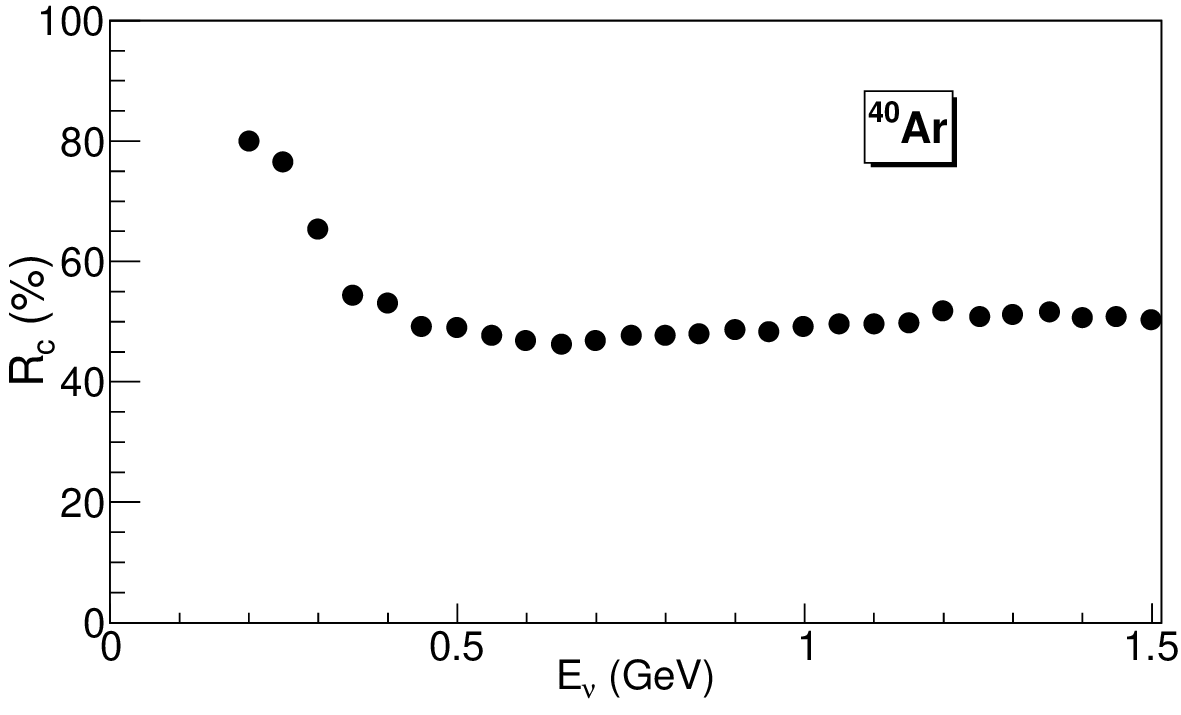}}
\subfigure[]{\includegraphics[scale=0.35]{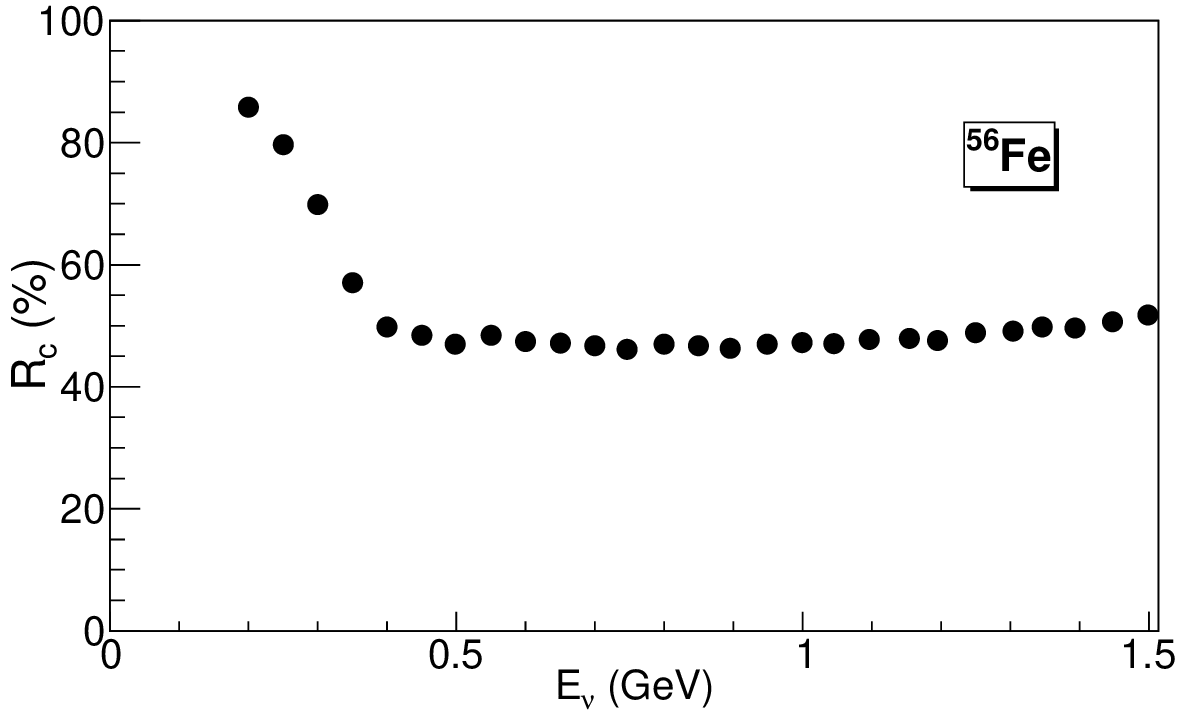}}
\subfigure[]{\includegraphics[scale=0.35]{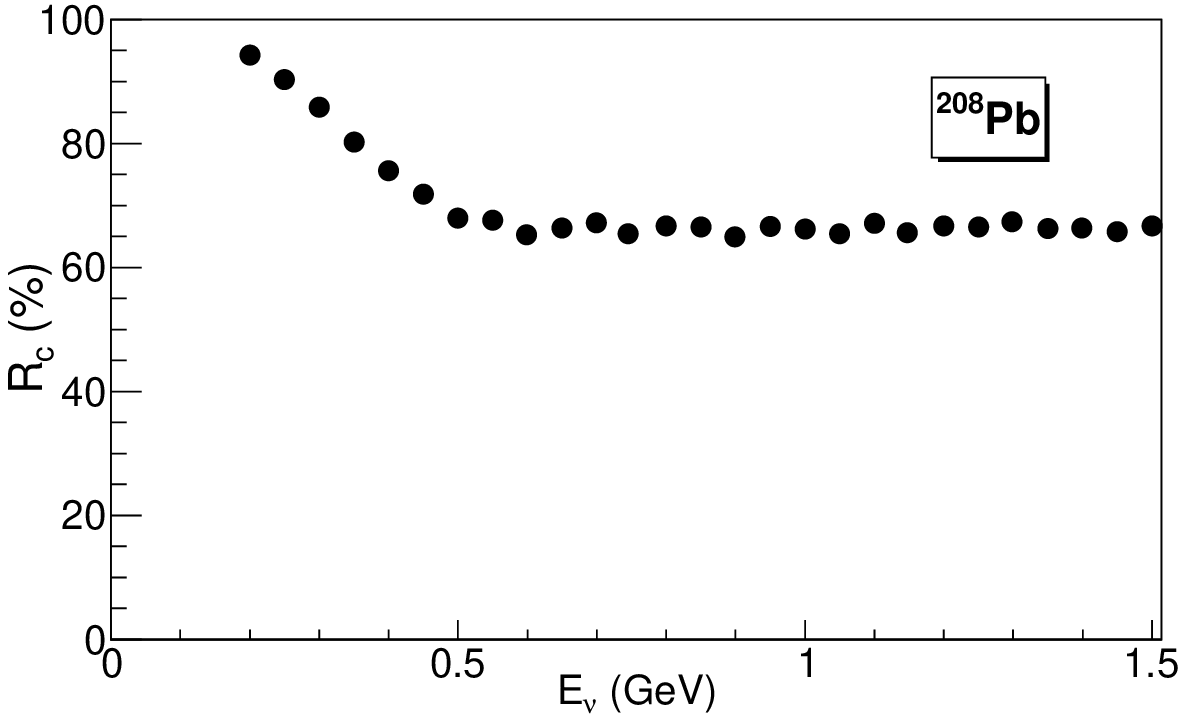}}
\caption{Percentage of false events, $R_c(i)$ according to Eq. (\ref{18}),
as function of neutrino energy for CCqe channels for the studied
target nuclei: a) $^{12}$C,  b) $^{16}$O, c) $^{27}${Al},
d) $^{40}${Ar}, e) $^{56}${Fe} and f) $^{208}${Pb}.
}
\label{figPercFalseEven} 
\end{figure*}

In Figure~\ref{figPercFalseEven} are shown the
percentage of false events, $R_c(i)$ according to Eq. (\ref{18}),
as a function of neutrino energy for CCqe channels for the studied
target nuclei: $^{12}$C,  $^{16}$O, $^{27}${Al},
$^{40}${Ar},  $^{56}${Fe} and  $^{208}${Pb}.
We note that the ratio is initializing in $250$ MeV with the higher
value ($\approx 80 \%$)
and then it goes to an averaged constant value.
The behavior is similar
to all the target nuclei  except for
$^{12}$C, mainly due that the Fermi gas model is not a good
description for light nuclei as carbon.  We can intuit
that the saturation effect of  $R_{c}(i)$
is because the neutrino has reached the maximum of interactions
within the space
of possible configurations of type $xp-xn$ created inside the nucleus
over 0.5 GeV.
This fact must be revised
when we will include in our simulation
more channels  coming from
resonances higher than $\Delta(1232)$.

\begin{table}[!ht]
\centering
\caption{
Percentage (\%) of fake events, $R_c(i)$
according to Eq. (\ref{18}),
when one uses the KM  and CIM methodologies
~\cite{Barb13,Mari11}. }
\label{falso negativo}
\begin{ruledtabular}
\begin{tabular}{ C{1cm} | C{0.7cm} | C{0.7cm} | C{0.7cm} | C{0.7cm} | C{0.7cm} | C{0.8cm} }
Channel& $^{12}$C & $^{16}$O & $^{27}${Al} & $^{40}${Ar}
& $^{56}${Fe} & $^{208}${Pb} \\\hline
A-KM    & 30 & 34 & 40  & 50  & 50  & 64  \\
A-CIM   & 30 & 34 & 38  & 48  & 48  & 66 \\ \hline
B-KM    & 50 & 58 & 68 & 76 & 82 & 90 \\
B-CIM   & 48 & 60 & 70 & 78 & 80 & 90 \\  \hline
C-KM    & 82 & 86 & 88 & 92 & 94 & 98 \\
C-CIM  & 84 & 88 & 90 & 92 & 94 & 96 \\ \hline
D-KM   & 62 & 70 & 74 & 78 & 80 & 88 \\
D-CIM  & 64 & 70 & 74 & 80 & 82 & 88 \\
\end{tabular}\end{ruledtabular}
\end{table}	

In Table~\ref{falso negativo} we present the results of fake events
obtained
with CRISP using KM and CIM formalism for the CCqe and CCres channels
in $^{12}$C, $^{16}$O, $^{27}${Al}, $^{40}${Ar}, $^{56}${Fe}
and $^{208}${Pb}.
The CIM model cross sections as a function of the neutrino
energy ~ \cite{Barb13}
were fitted to a fourth degree polynomial  to include in CRISP.
In the first column of Table~\ref{falso negativo}, we shown the
channels interaction labeled as in Table I. The next columns shown
the evolution of percentage of
fake events as increasing mass number according to
the target nuclei.
The table can go through a
solid  nucleus mass analyzing
the contribution for each channel.
The inputs
of the A channel are lower for all the nuclei, following
in ascending order D and B.  The maximum is obtained
from C reaction
($\nu_{\mu}+n\rightarrow \mu^{-}+\Delta^{+}\rightarrow \mu^{-}+\pi^{+}+n$)
for all the target nuclei being in average $\approx 90 $ \%.
On the other point of view, relative to the channel reaction,
we note
that the fake events increase as well the mass increases,
being minimal in carbon and maximum in the lead.
Summarizing, we can observe that with the
growth in the atomic number
and atomic mass of the target nucleus, which increases
the percentage of false events
due to the appearance of the nuclear
structure effects and
the interactions among the several nucleons.

The employment of different formalisms
like KM and CIM  for the neutrino cross section of primary interaction
has almost any effect on the percentage
of false events because they are a direct consequence
of the intranuclear cascade and not of primary exchanges.

\subsection{
Energy distribution of the emitted pions and muons }

The pion spectrum
in the CRISP model  is calculated as
\[
\frac{d\sigma}{dT_\pi}=\sigma_g \frac{N_{ev}(T_\pi)}{N_0 \Delta T_\pi},
\]
a
where $N_{ev}(T_\pi)$ is the number of events
for a specific channel producing
a pion with a given isospin with energy
between $T_\pi$ and $T_\pi+\Delta E_\pi$.
%
\begin{figure*}[t]
\centering
\subfigure[]{\includegraphics[scale=0.42]{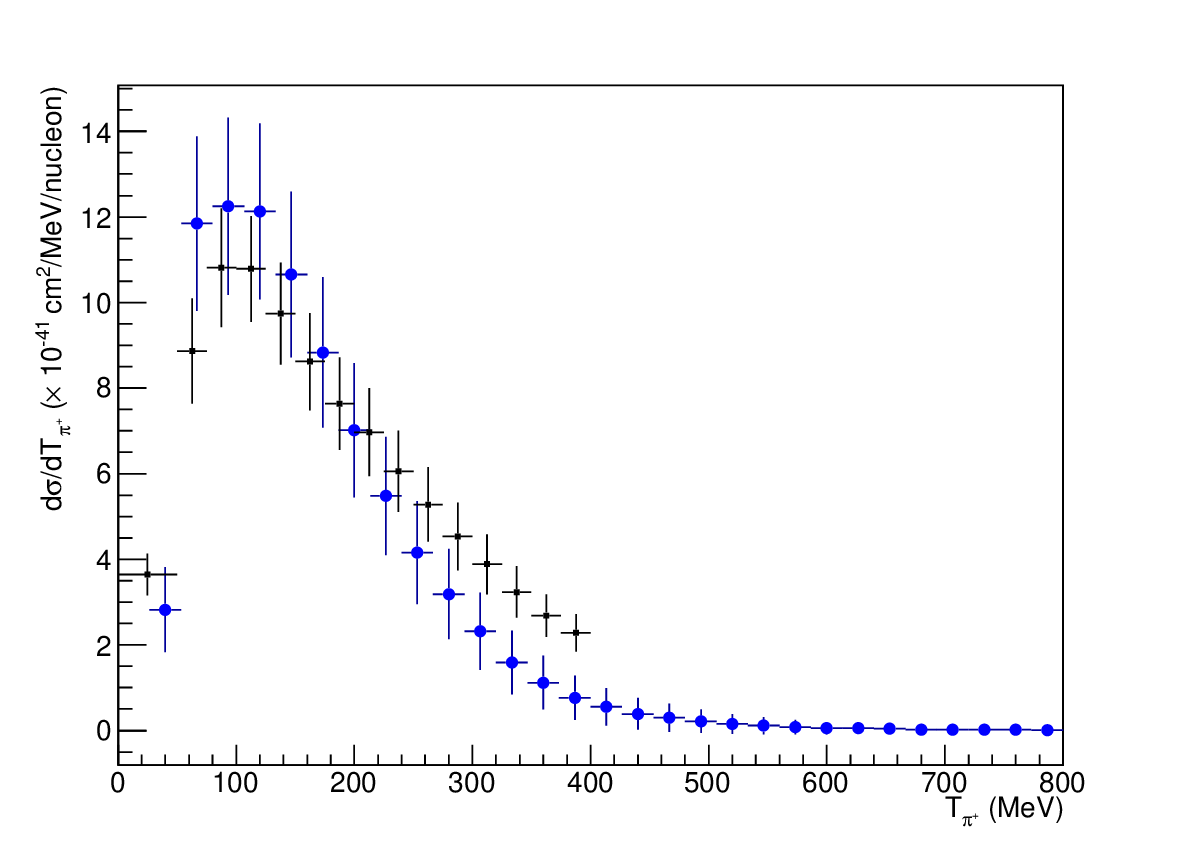} \label{figEnergDist:pion2}}
\subfigure[]{\includegraphics[scale=0.42]{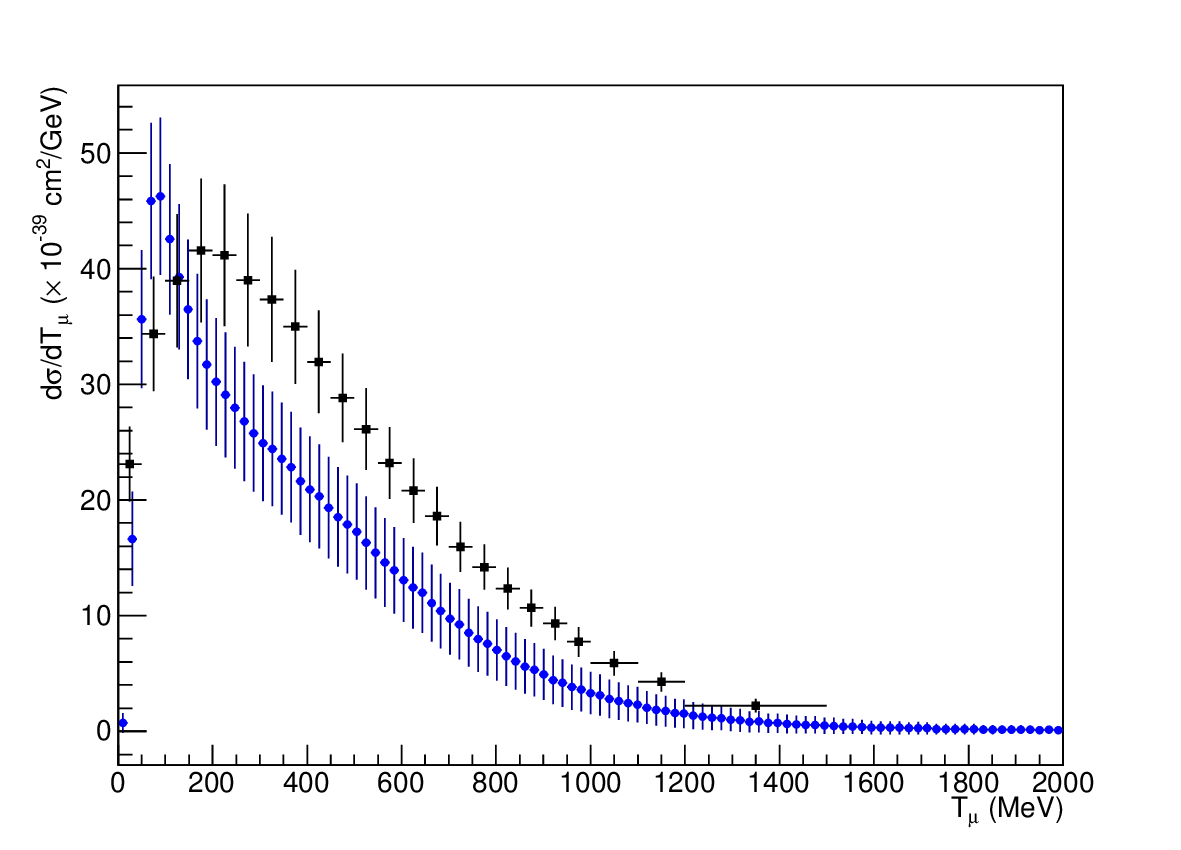} \label{figEnergDist:muon2}}
\caption{\label{pion-muon-dist}Energy distribution for  $\pi^+$  (left panel)
 and  $\mu^-$ (right panel),
for reaction induced with $\nu_\mu$ with the energy
of $\sim 1$ GeV on $^{12}$C. Experimental data (black cross)
of MiniBooNE~\cite{MiniBooNE10,MiniBooNE11} are also shown
for comparison.
The error bars in the theoretical calculations
are statistical due the propagation of error in the Monte Carlo and from
the deviation in the neutrino flux.
}			
\label{figEnergDist} 
\end{figure*}
The MiniBooNE experiment measured the positive pion spectrum
for $E_\nu \sim 1$ GeV on $^{12}$C~\cite{MiniBooNE10,MiniBooNE11}.
In the left panel of Figure~\ref{pion-muon-dist}, we show our calculation
 averaged over the published MiniBooNE flux
for $\pi^+$ \cite{MiniBooNE09}
in comparison with the experimental data. We observe that both
calculation and data show a similar shape with the
peak around 80 MeV and a large tail at high energies.
Quantitatively there is a good agreement between calculation
and experiment, notably in the peak region. At energies
above 250 MeV, the calculation underestimates the experimental data.
It is likely that this effect is related to that
we were not included higher mass resonances in the present calculations.

Similarly,  the ejected muon distribution is calculated  as a function
of the kinetic muon energy as
\[
\frac{d\sigma}{dT_\mu}=\sigma_g \frac{N_{ev}(T_\mu)}{N_0 \Delta T_\mu}.
\]
The right panel of Figure~\ref{pion-muon-dist} shows the muon distribution
calculation as a function of the kinetic muon energy,  averaged over the published
MiniBonNE flux for $\pi^+$ \cite{MiniBooNE09},
in comparison with the experimental MiniBooNE data for $E_\nu \sim 1$ GeV.
Here, we note that our theoretical calculation
is slightly lower than
the experimental results, but the behavior and the peak position
are in a well agreement with data.
Relative to these calculations:
(i) we do not adjust the pion mass resonances to reproduce the
experimental spectra as it was done by Lalakulich
and Mosel \etal in \cite{Lalakulich2013};
(ii) we are taken into account only the contribution of
the delta resonance, for this reason, we do not implement
other reaction channels in the neutrino generator with
other resonances, presented in the intranuclear cascade in
these calculations and;
(iii) our formalism is not including an angular distribution for
the ejected particle.
Then, our model will not be in completely agreement with some experimental data,
in comparison with another model performed by such issue.
Also, it is important to remember that we use a Fermi gas model,
which is not the best choice, especially for $^{12}$C,
so that structure does not have any physical significance.

Some final words are devoted to the comparison with another theoretical
model, as such that performed in Ref. ~\cite{Lalakulich2013}.
The analysis performed here is in many aspects similar to
the one presented in Ref.~\cite{Lalakulich2013},
where medium's effects on neutrino nucleus interaction were studied.
The most relevant differences between the approach used here and
that in Ref.~\cite{Lalakulich2013} are related to the modeling
of the bound nucleon dynamics. A summary of these differences is
the following:

\begin{enumerate}
\item
In the CRISP model the nucleus is described as a global Fermi gas,
while in Ref.~\cite{Lalakulich2013} is used a local
Thomas-Fermi approach.
\item
As a consequence of the first difference, in CRISP model
the Pauli blocking mechanism is accounted for strictly, while
in Ref.~\cite{Lalakulich2013} it is considered statistically.
Careful analysis of the advantages of a strict Pauli blocking
mechanism are presented in Refs.~\cite{Deppman2004, Rodrigues2004}.
\item
With the inclusion of Fermi motion and
rigorous Pauli blocking, some nuclear effects emerge naturally
in the calculations with the CRISP model,
such as shadowing effect, that is present in photoabsorption
and in meson production, for example in Refs.~\cite{Deppman06,Israel2011b}.
Also, medium's effects on resonance propagation are naturally
accounted in the CRISP model.
\end{enumerate} 

These differences are relatively more important for energies
near the reaction threshold, and should practically disappear
as the incoming particle energy increases. At first sight,
the aspects mentioned above could explain why the CRISP model
gives better results as compared to experimental data
than the calculations in Ref.~\cite{Lalakulich2013},
however the disagreement between the both calculations
seems to be too large to be attributed only to those
different methods used in each model.

The medium's modifications in $\Delta$ resonance, for instance,
were first observed in photoabsorption measurements
and were mainly attributed to Fermi motion and Pauli blocking effects
\cite{photoab1, photoab2, photoab3},
although some effects from the coherent sum of resonant
and direct channels could be
observed \cite{Hirata}.
In Ref.~\cite{Lalakulich2013} the authors inform
to have included the $\Delta$ resonance broadening through
the Salcedo and Oset model \cite{SalcedoOset},
but in their spectral function are also encompassed
for the bound nucleon, both
Fermi motion and Pauli blocking effects.
It is possible, then, that the $\Delta$ resonance broadening
is taken into account twice: one time in the modeling of the
resonance, and the another time by the nuclear effects
already considered in their nuclear model.
This  way of counting
could explain why their calculation underestimates the
cross section in the resonance peak energy, since the broadening
of the resonance width results in a reduction of the cross section
at the peak. Also, it can explain the shift of the peak energy
to lower energies, since the combination of Fermi motion and
Pauli blocking produces such effect.

\section{Conclusion \label{Conc}}

In the present work we report an extensive analysis of nuclear effects
in neutrino-nucleus interaction. For this purpose, a simple model of
neutrino-nucleon, which was called Kinetic Model, is used together
with the CRISP model to take into account the nuclear effects. This
simple model has three free parameters for all channels analyzed
in this study.
We determined these parameters by fittings to the
specific channels to neutrino-deuterium experimental data.

The calculations were performed for neutrino energies from 0.2
to 1.5 GeV for $^{12}$C, $^{16}$O, $^{27}${Al}, $^{40}${Ar}, $^{56}${Fe}
and $^{208}${Pb}. We calculated the cross section for all nuclei in
the whole energy range using CCqe and CCres channels. Where
data is available, a comparison between calculation and experiment
was provided. The pion and muon spectra are also calculated
and compared to the experimental data showing a fair agreement.

For the set of target nuclei employed, we performed
an exploratory study of the fake events generated in several reactions.
This study has shown
that the percentage of CCqe fake events for $^{12}$C, important
for MiniBooNE, are in the same order of  $\approx 30$ \% of previous
works~\cite{Lalakulich2013,Eric2016}. Whereas than for other nuclei
the percentage of fake events increases as well the nuclei masses
increases due to the structure effect and multinucleon excitations in
the nucleus. Using two different formalisms of neutrino-nucleon cross
sections was shown than the percentage of fake events are almost
independent  of the primary interactions because they are a
direct consequence of intranuclear cascade.
In a future work, we will improve the simple Kinetical Model
used for the primary interaction using the Consistent
Isobar Model-CIM~\cite{Barb13,Mari11}.

Our conclusions are that nuclear effects are
decisive
for understanding the neutrino-nucleus interaction, and the most
substantial effect is the interaction of the produced particles with the nucleus.
Another important
effect that appears mainly for neutrino energies below $\sim 0.5$ GeV
are the nuclear binding energy, Fermi motion and Pauli blocking.
For all studied channels, we observed the
subthreshold reaction.
Finally, we predict that nuclear structure
plays a relevant role in this energy range.

\begin{acknowledgments}
Part of this work was
performed
in the frame of the academic cooperation
agreement between UESC and InSTEC. O. Rodr\'iguez and F. Guzm\'an
would like to recognize the provisions of UESC for the conclusion
of this work. D. Vargas,  A. R. Samana and F. Velasco
thank the financial support of
Funda\c{c}\~ao de Amparo \`a Pesquisa do Estado da Bahia (FAPESB)
and CAPES-AUXPE-FAPESB-3336/2014/Processo no: 23038.007210/2014-19.
A. Deppman thanks the financial support of CNPq/305639/2010-2.
C.Barbero and A. Mariano are fellows of the CONICET, CCT La Plata
(Argentina) and thank for partial support under Grant PIP No 0349.
We sincerely thank Professor Juan J. Godina-Nava for his
very careful and judicious reading  of the manuscript.

Finally, the authors thank the financial support of Brazilian
agencies CNPq and CAPES.

\end{acknowledgments}


%
\end{document}